\documentclass[11pt]{article}
\usepackage{amsmath,bm,bbm}
\usepackage{geometry}
\usepackage{hyperref}

\def\gtwid{\mathrel{\raise.3ex\hbox{$>$\kern-.75em\lower1ex\hbox{$\sim
$}}}}
\def\vio{\mathrel{\hbox{$R$\kern-.60em\hbox{$/
$}}}}
\usepackage{graphicx}
\usepackage{amssymb}
\def\lsim{\mathrel{\raise.3ex\hbox{$<$\kern-.75em\lower1ex\hbox{$\sim$}}}}
\def\gsim{\mathrel{\raise.3ex\hbox{$>$\kern-.75em\lower1ex\hbox{$\sim$}}}}

\newcommand{\gev}{\text{GeV}}

\newcommand{\pb}{\text{pb}}

\newcommand{\be}{\begin{equation}}
\newcommand{\ee}{\end{equation}}
\newcommand{\bea}{\begin{eqnarray}}
\newcommand{\eea}{\end{eqnarray}}

\begin{document}

\title
{Fermion WIMPless Dark Matter at DeepCore and IceCube}
\author{Vernon Barger$^{1}$, Jason Kumar$^{2}$, Danny Marfatia$^{3}$ and Enrico Maria Sessolo$^{3}$\\[2ex]
\small\it ${}^{1}$Department of Physics, University of Wisconsin, Madison, WI 53706, U.S.A.\\
\small\it ${}^{2}$Department of Physics and Astronomy, University of
Hawai'i, Honolulu, HI 96822, U.S.A.\\
\small\it ${}^{3}$Department of Physics and Astronomy, University of
Kansas, Lawrence, KS 66045, U.S.A.}

\date{}

\maketitle

\begin{abstract}
We investigate the prospects for indirect detection of fermion WIMPless dark matter
at the neutrino telescopes IceCube and DeepCore. The dark matter annihilating in the Sun is
a hidden sector Majorana fermion that couples through Yukawa couplings to a connector particle
and a visible sector particle, and it exhibits only spin-dependent scattering with nuclei via couplings
to first generation quarks. We
consider cases where the annihilation products are taus, staus, or sneutrinos of the three generations.
To evaluate the muon fluxes incident at the detector, we propagate the neutrino spectra through the
solar medium and to the Earth and account for the effects of neutrino oscillations, energy losses due
to neutral- and charged-current interactions, and tau regeneration. We find that for the stau and
sneutrino channels, a 5 yr 3$\sigma$ detection of dark matter lighter than about 300~GeV is possible
at IceCube for large Yukawa couplings or for dark matter and connector particles with similar masses. The
tau channel offers far better detection prospects. However, due to its lower energy threshold and better
muon background rejection capability, DeepCore is able to detect signals in all annihilation channels and
for a wider range of dark matter masses.
\end{abstract}


\newpage

\section{Introduction}

There has long been interest in the use of neutrino experiments
for indirect dark matter (DM) detection, with a particular focus on neutrino signals from
DM annihilation in the Sun. The basic idea is that DM particles
can be captured by the Sun if they lose sufficient
kinetic energy through elastic scattering from solar nuclei.  This
leads to an enhanced DM number density in the solar core,
where DM can annihilate to standard model (SM) products. These
products in turn decay, producing a neutrino flux which can be
detected by Earth-based detectors.
Because the Sun is (for most models
under consideration) in equilibrium, the DM capture rate
is equal to twice the annihilation rate.  Thus, the neutrino flux on Earth
is determined by the capture rate, which in turn is determined (up to a
few ${\cal O}(1)$ factors related to solar physics) by the DM
mass $m_{\chi}$ and the DM-nucleon scattering cross section $\sigma$. This is one of
the key advantages of neutrino-based indirect searches for DM: the
event rate is directly related to the scattering cross section, without
many of the astrophysical uncertainties which attend other indirect detection
strategies.  Moreover, it is in principle possible to directly compare search results
at neutrino experiments with those at direct-detection experiments, which also
measure the DM-nucleon scattering cross section; for a recent discussion, see \cite{Wikstrom:2009kw}.  This allows one
to corroborate a signal at one type of experiment with a signal at a very
different type of experiment, which is desirable.

The downside, however, is that for much of the $(m_\chi,\sigma)$-parameter
space, sensitivities for current and future neutrino detectors have
already been surpassed by direct-detection bounds, implying that
neutrino detectors may have difficulty providing new input to DM
studies.  However, there are two scenarios where
neutrino experiments are expected to shine. One is at low-mass
($m_\chi \sim 4-10~\gev$).  In this mass range, bounds from direct-detection experiments tend to become significantly worse because the
nuclear recoil energies often fall below the experimental threshold.
Experiments such as Super-Kamiokande (Super-K) and other proposed water Cherenkov or
liquid scintillator-type detectors can provide the best bounds in this
range~\cite{lightDMdetect}, and could be sensitive to spin-independent (SI) scattering cross sections
$\sigma_{SI} \sim 10^{-5}~\pb$. DM in this
range can potentially explain the annual modulation signal seen by the
DAMA experiment (\cite{lightDMDama}, though see also \cite{Fairbairn:2008gz}), as
well as unexplained events recently reported by the CDMS and
CoGeNT Collaborations~\cite{Ahmed:2009zw,Aalseth:2010vx,Cogentcdmstheory}.
There is thus considerable interest to cross-check these results
with neutrino detectors.

Another scenario for which neutrino experiments can
excel is for models where the DM-nucleon scattering is largely
spin-dependent (SD), rather than SI~\cite{sigmaSDneutrino}.
The reason is that DM can then be captured
by SD scattering from hydrogen in the Sun, leading to
significant neutrino fluxes.  Current direct-detection experiments provide
much less sensitivity to SD scattering and the best bounds
are set by indirect detection experiments, even at large $m_\chi$.  This has
been the focus of large neutrino detectors, such as AMANDA or
IceCube.  For example, with 1800 live days of running, IceCube could be sensitive
to ${\cal O} (100~{\rm GeV})$
DM with
$\sigma_{SD} \sim 10^{-5}~{\rm pb}$~\cite{Braun:2009fr}.
Although the relatively high threshold energy of these detectors limits their utility in
studying the low-mass DM, their large size makes them ideal for studying
higher-mass DM ($m_{\chi} \geq 100~{\rm GeV}$) with $\sigma_{SD} \gg \sigma_{SI}$.

Direct detection
is much more sensitive to $\sigma_{SI}$ due to the possibility of coherent
scattering in the nucleus.  Direct-detection experiments, including 
CDMS and XENON100, currently have a sensitivity to $\sigma_{SI}$ in the 
${\cal O} (100~{\rm GeV})$ mass range of 
about 3 orders of magnitude~\cite{Ahmed:2009zw,XENON100limits} greater than
the IceCube projected 1800 d sensitivity to $\sigma_{SD}$.
But for models with $\sigma_{SI} \ll \sigma_{SD}$, complementary coverage from 
neutrino detectors becomes important.  This complementary coverage is especially 
important  if one goes beyond models of weakly interacting massive particles (WIMPs).
The recently proposed WIMPless model of dark matter~\cite{Feng:2008ya,Feng:2008dz} is
notable because it provides a very generic hidden sector
DM candidate which naturally has about the right
relic density to match astronomical observations, regardless of
the mass of the DM particle or the details of the hidden
sector.  This versatility suggests the possibility of specific
WIMPless models with almost exclusively SD scattering,
for which the best detection prospects would lie at neutrino
detectors.

We present a model of WIMPless dark matter
where the DM candidate is a Majorana fermion in
the hidden sector.  At tree level, it will exhibit only
SD scattering with SM nuclei.  We find that significant neutrino fluxes on Earth arise from
models where DM annihilates either to tau,
stau, or sneutrino pairs.  We consider the detection
prospects for this type of model at DeepCore and IceCube.

The paper is organized as follows: In Sec.~2, we review WIMPless dark matter and
describe the properties of the fermion candidate DM particle.
In Sec.~3, we obtain the neutrino spectra arising from the decay of various annihilation products.
In Sec.~4, we discuss the predicted event rates at DeepCore and IceCube, and we present our
conclusions in Sec.~5.

\section{WIMPless Dark Matter}

Here we briefly review the WIMPless DM scenario~\cite{Feng:2008ya}.
This scenario is closely related to gauge-mediated supersymmetry
breaking (GMSB)~\cite{Dine:1993yw}. Our theory consists of a supersymmetry (SUSY)-breaking sector,
the visible minimal supersymmetric standard model
(MSSM) sector and a hidden sector, plus a \textit{connector} sector which we introduce in the next
subsection. We assume there is at least one chiral superfield $S$ in the SUSY-breaking sector, and
there are messenger fields
$\Phi$, $\bar{\Phi}$ between the SUSY-breaking and MSSM sectors and $\Phi_\chi$, $\bar{\Phi}_\chi$ between the
SUSY-breaking and hidden sector. $S$ is coupled to the messenger fields through Yukawa couplings
in the superpotential $W=\lambda\bar{\Phi}S\Phi + \lambda_\chi \bar{\Phi}_\chi S\Phi_\chi$.
The effect of GMSB on the MSSM sector is well-known: the chiral field $S$ acquires
a vacuum expectation value $\langle S\rangle=M+\theta^2 F$.  The Yukawa couplings generate
messenger mass terms of order $m_{mess}\sim\lambda M$ and
messenger mass splittings of order $F_{mess}=\lambda F$. Once the heavy gauge
messengers are integrated out, the low-energy effective theory has
a new SUSY-breaking soft scale $m_{soft}$ which is generated by
diagrams with the messengers running in the loops,
\begin{equation}
m_{soft} \sim {g^2 \over (4\pi )^2} \left( {F_{mess} \over m_{mess}} \right)={g^2 \over (4\pi )^2}
{F \over M}\,,\label{msoft}
\end{equation}
where $g$ is the largest relevant gauge coupling. The effect in the
hidden sector is qualitatively the same, and
we assume that some unbroken symmetry stabilizes a particle in
the hidden sector at the soft SUSY-breaking scale, whatever it may be.
The soft SUSY-breaking scale in the hidden sector is then given by
\be
m_{\chi} \sim {g_{\chi}^2 \over (4\pi )^2} \left( {F_{mess\chi}
\over m_{mess\chi}} \right)={g_\chi^2 \over (4\pi )^2} {F \over M}\,,\label{mchi}
\ee
where $F_{mess\chi}=\lambda_{\chi} F$ and $m_{mess\chi}\sim\lambda_{\chi} M$.
Because the ratio $F/M$ is determined by the dynamics
of the SUSY-breaking sector, it appears in the same way in the soft
scale in both the hidden sector and MSSM sector.  We thus find
\be
{g_\chi^4 \over m_\chi^2} \sim {g^4 \over m_{soft}^2}
\propto \left( {M \over F} \right)^2.
\label{sigmascaling}
\ee
The ratio $g^4/m^2$ sets the annihilation cross section
for a particle of mass of order $m$ through gauge interactions of strength $g$, and
the cross section in turn determines the thermal relic density,
($\Omega \propto \langle \sigma v \rangle^{-1}$~\cite{Zeldovich:1965}).
To obtain a thermal relic density which matches astronomical
observations, a DM candidate would need an annihilation
cross section $\langle \sigma v \rangle \sim 1~\pb$.  The
``WIMP miracle'' is the amazing fact that, when calculated at the
electroweak scale, the ratio $g_{weak}^4 /m_{weak}^2$
yields an annihilation cross section which is in fact close to
$1~\pb$.  But we see from Eq.~(\ref{sigmascaling}) that the annihilation
cross section for a stable particle at the hidden sector soft
SUSY-breaking scale will be approximately the same; this DM
candidate will thus automatically have approximately the right
relic density.  Most importantly,
the scale $m_\chi$ of the DM is a free parameter, thus motivating DM
searches over a wide range of masses.

\subsection{Fermionic Model}

In the WIMPless scenario, DM is stabilized by a hidden sector
symmetry (which may be discrete).
We let $\hat{X}_{L,R}$ be chiral supermultiplets in the hidden
sector, with a fermionic mass eigenstate of $X_{L,R}$ (henceforth denoted as $X$) being the
lightest particle charged under the stabilizing symmetry.
WIMPless DM interacts with the SM through
Yukawa couplings, which permit the exchange of exotic ``connector"
particles.  Consider the interaction superpotential,
\bea
W &=& \lambda_{Li} \hat{X}_L \hat{Y}_{L} \hat{Q}_{Li} + \lambda_{Ri} \hat{X}_R \hat{Y}_{R} \hat{Q}_{Ri}
+m_{Y} \hat{Y}_{L} \hat{Y}_{R}
\nonumber\\
&\,& +\lambda'_{Lj} \hat{X}_L \hat{Y}_{L}^{lep.} \hat{L}_{Lj} 
+ \lambda'_{Rj} \hat{X}_R \hat{Y}_{R}^{lep.} \hat{L}_{Rj}
+m_{Y^{lep.}} \hat{Y}_{L}^{lep.} \hat{Y}_{R}^{lep.},
\eea
where $\hat{Q}_{Li,Ri}$ are MSSM chiral quark multiplets,
$\hat{L}_{Li,Ri}$ are MSSM chiral lepton multiplets, and
$\hat{Y}_{L,R}$ and $\hat{Y}_{L,R}^{lep.}$ are the exotic connector particles which are
charged under both the MSSM and the hidden sector
symmetry. Gauge-invariance requires the $\hat{Y}_{L,R}$ to be
chiral under the MSSM, so their fermion components behave like exotic fourth generation quarks
(similarly, the fermions of $\hat{Y}_{L,R}^{lep.}$ are exotic fourth generation leptons).
The mass terms $m_{Y}$, $m_{Y^{lep.}}$ are thus set by electroweak symmetry breaking,
and are determined by the Higgs vacuum expectation value and Yukawa couplings.

The mass of the exotic quarks is constrained by perturbativity and
precision electroweak data to the range
$m_Y \lesssim  600\,{\rm GeV}$~\cite{Kribs:2007nz,4thgendirectsearch,Alwall:2010jc}.
Since the DM particle $X$ is the lightest particle
charged under the stabilizing hidden sector symmetry, it is constrained to
be lighter than any of the exotic connectors.

Since the exotic squarks $\tilde Y_{L,R}$ can get mass terms which are decoupled from
electroweak symmetry breaking, there is no relevant upper bound on $m_{{\tilde Y}_{L,R}}$.
But $m_{{\tilde Y}_{L,R}}$ can be bounded from below by direct searches
at colliders.  The exotic squarks can be pair produced through QCD
processes, with each decaying via ${\tilde Y}_{L,R} \rightarrow X + u,d$.  The signal at a hadron
collider would be exclusive dijet production with missing transverse
energy, which is the same signature as that of leptoquark pair production, with decay
to a quark and neutrino.  A CDF Collaboration search for this
signature~\cite{Aaltonen:2009xp} places
a bound on the leptoquark mass at $m_{LQ} > 187~\rm{GeV}$.  Since, in our case, the
missing energy arises from a relatively heavy-DM particle $X$ (as opposed to a
nearly massless neutrino), the lower bound on the mass of the exotic squark
will be even weaker, and for our analysis it will not be constraining.

The DM candidate $X$ can be either a Dirac or Majorana fermion.  We thus see that DM-nucleon
scattering can arise from $s$- or $u$-channel exchange of
the $Y$ multiplets, as represented in Fig.~1.
\begin{figure}[ht!]
 \begin{center}
  \includegraphics[width=90mm, height=45mm]{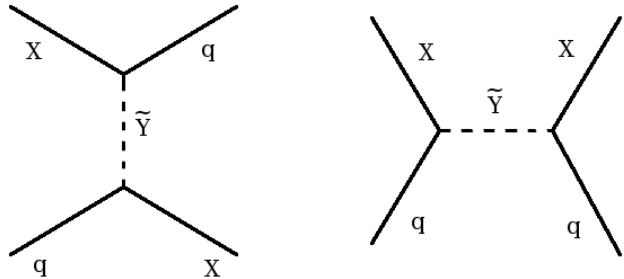}
     \caption{The $u$- and $s$-channel Feynman diagrams for $Xq$ scattering.}
   \end{center}
\end{figure}
After a Fierz transformation, we can
write the DM-parton scattering matrix element in
terms of an effective operator,
\bea
{\cal O} &=&
{\lambda_{Li}^2 \over 2(m_X^2 -m_{{\tilde Y}_L}^2)^2} (\bar q_i \gamma^{\mu} P_L q_i)
(\bar X \gamma_{\mu} P_R X)
+{\lambda_{Ri}^2 \over 2(m_X^2 -m_{{\tilde Y}_R}^2)^2} (\bar q_i \gamma^{\mu} P_R q_i)
(\bar X \gamma_{\mu} P_L X)\,.
\eea
We assume from here on that the scalars from the $Y_L$ and $Y_R$
multiplets have degenerate mass; $m_{{\tilde Y}_{L}} =m_{{\tilde Y}_{R}}= m_{\tilde Y}$.
If we only retain terms which provide a velocity-independent
contribution to DM-nucleon scattering, we can write
this operator as
\bea
{\cal O} &\approx& {\lambda_{Li}^2 +\lambda_{Ri}^2 \over 8(m_{\tilde Y}^2 -m_X^2)}
[(\bar q_i \gamma^{\mu} \gamma_5 q_i )
(\bar X \gamma_{\mu} \gamma_5 X)
-(\bar q_i \gamma^{\mu} q_i )
(\bar X \gamma_{\mu} X)]\,.
\eea
We find that if $X$ is a
Dirac fermion, then
\bea
\sigma_{SI} &=&  {1\over 4}
{m_r^2 \over 64\pi (m_{\tilde Y}^2 - m_X^2)^2}
\left[\sum_i (\lambda_{Li}^2 +\lambda_{Ri}^2 )
(Z B^p_i+(A-Z)B^n_i )\right]^2
\nonumber\\
\sigma_{SD} &=& {1\over 4}
{m_r^2 \over 4\pi (m_{\tilde Y}^2 - m_X^2)^2}
{J+1 \over J}\left[\sum_i (\lambda_{Li}^2 +\lambda_{Ri}^2 )
(\langle S_p \rangle \Delta_i^{(p)} + \langle S_n \rangle \Delta_i^{(n)})\right]^2,
\eea
where the $1/4$ factor accounts for the fact that
the DM is not its own antiparticle, so only the $s$- or $u$-channel
will contribute to $Xq \rightarrow Xq$ scattering. $m_{r}=m_{X}M_{N}/(m_{X}+M_{N})$ is
the reduced mass of the DM-nucleus system.
The vector-vector part of the effective operator generates $\sigma_{SI}$, while the axial-axial
part generates $\sigma_{SD}$.
The $B_i^{(p,n)}$ parametrize the quark content of the nucleon, and the $\Delta_i^{(p,n)}$ its
quark spin content. The $\langle S_{p(n)}\rangle$ are the expectation values of the spin content
of the proton~(neutron) group in the nucleus. The numerical values of $\Delta_{i}^{p}$ corresponding
to structure functions including next-to-next-to-leading order QCD corrections are~\cite{Mallot:1999qb},
\begin{equation}
\Delta_u^p = 0.78\pm0.02\,,\,\,\,\,\,\,\,\Delta_d^p = -0.48\pm0.02\,,\,\,\,\,\,\,\,\Delta_s^p 
= -0.15\pm0.02\,.\label{Deltas}
\end{equation}
As noted in Ref.~\cite{Ellis:2008hf}, uncertainties in quark spin contents can lead to substantial 
variations in $\sigma_{SD}$.

If the DM is Dirac, then a $\sigma_{SD}$
which is potentially detectable at DeepCore would already
be ruled out by direct-detection constraints on $\sigma_{SI}.$ We therefore focus on the case where $X$ is a
Majorana fermion which is stabilized by a $Z_2$ symmetry (so $CXC = X$).
The vector-vector part of the effective operator is then identically
zero and the scattering cross section is necessarily SD
($\sigma_{SI}=0$), and we find
\bea
\sigma_{SD} &=&
{m_r^2 \over 4\pi (m_{\tilde Y}^2 - m_X^2)^2}
{J+1 \over J}
\left[\sum_i (\lambda_{Li}^2 +\lambda_{Ri}^2 )
(\langle S_p \rangle \Delta_i^{(p)} + \langle S_n \rangle \Delta_i^{(n)})
\right]^2\,,\label{MajSD}
\eea
since now both $s$ and $u$~channels contribute.  Note that we have
only included the velocity-independent terms, and only at tree level.
Since the bounds on $\sigma_{SI}$ are $\sim 4$ orders of magnitude
stronger than the expected IceCube bounds on $\sigma_{SD}$, one might
wonder if this model could be probed by direct-detection
experiments via velocity-dependent or loop-induced spin-independent
effective operators. One would expect velocity-dependent operators
to be suppressed by at least~$10^{-6}$. Also, loop-induced
SI effective operators would be suppressed by \mbox{$(\alpha /4\pi)^2 \sim 10^{-6}$}. Thus, with
further improvements, direct-detection probes of $\sigma_{SI}$ induced by
one-loop diagrams may be comparable to probes of $\sigma_{SD}$ at tree level
by IceCube.  This is an interesting subject for future study, but is beyond
the scope of the present work.

In our model, DM can have axial-vector couplings to the light
and/or the heavy quarks.  Scalar DM coupling to heavy quarks was
studied in detail in~\cite{Shifman:1978zn}, but no such study has been
performed in the case of axial-vector coupling.  Instead, we consider
the case where DM couples to light quarks; if DM
couples to more than one quark generation, then there is a potential for
dangerous flavor-changing neutral currents.  As a
simple solution to this constraint, we assume that the only
DM-quark coupling is to the first generation. We take
$\lambda\equiv\lambda_{Lu}=\lambda_{Ru}=\lambda_{Ld}=\lambda_{Rd}$, and $\lambda_{Ls}=\lambda_{Rs}=0$.

\subsection{Annihilation}

In WIMPless DM models, the annihilation cross section to
hidden sector particles is naturally $\langle \sigma_{ann.} v \rangle
\sim 1~\pb$ at the time of decoupling.
From the form of the
Yukawa couplings, we see that DM can also annihilate to MSSM
matter multiplets.
The annihilation cross section to MSSM particles should not be very much
larger than $1~\pb$, or else annihilation will dilute the relic density and
ruin the WIMPless miracle.  But in order to get a significant flux at neutrino
detectors, it is necessary for the annihilation branching fraction to
MSSM particles to be comparable to unity.

Since we are considering the case where the DM is Majorana, annihilation to
light fermions is chirality/$p$-wave suppressed. Also, in order to obtain SD scattering
without inducing dangerous flavor-changing neutral currents, we assume that the only quark multiplets
which the DM couples to are the first generation multiplets.
So DM annihilation to quarks will be suppressed; if the DM
annihilation to the hidden sector is not suppressed at current times, then the
branching fraction for DM in the Sun to annihilate to quarks will be very
small.  One caveat to this statement is that if the light hidden sector particles
are also fermions, then annihilation to the hidden sector is also chirality/$p$-wave
suppressed.  This does not significantly suppress the annihilation cross section
at decoupling, but may suppress the annihilation cross section to hidden sector
particles in the Sun by enough to allow the branching fraction to quarks to be
sizable.  In this case, DM annihilation to quarks in the Sun can be
a significant source of neutrinos observed at IceCube.
\begin{figure}[ht!]
 \begin{center}
  \includegraphics[width=90mm, height=45mm]{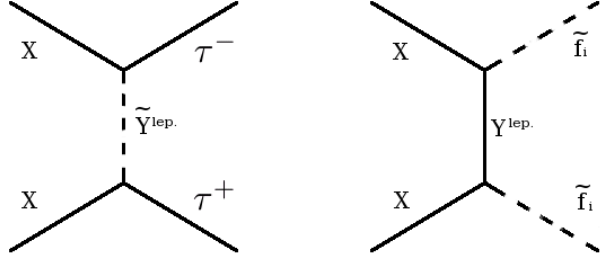}
     \caption{Feynman diagrams for DM annihilation
     into $\tau^-\tau^+$ and into $\tilde{f}_{i}\bar{\tilde{f}}_{i}$.}
   \end{center}
\end{figure}
But if there are light vectors or scalars (perhaps Goldstone bosons) in the
hidden sector, then annihilation to these products is not suppressed, and
the cross section for hidden sector annihilation would be expected to be
$\sim 1\,{\rm pb}$.  To get an annihilation cross section to SM particles
of comparable magnitude, one must either consider squark final states, or
make use of Yukawa couplings to SM leptons. Then, the only relevant MSSM
annihilation products are $\tau$, ${\tilde l}_{L,R}$ and
the first generation squarks ${\tilde u}_{L,R}$ and ${\tilde d}_{L,R}$.
These annihilation cross sections are given by
\bea
\sigma_{XX\rightarrow \tau \bar \tau} v &=&
{(\lambda'^2_{L\tau} +\lambda'^2_{R\tau })^2  m_{\tau}^2 \over
32\pi (m_{\tilde Y^{lep.}}^2 + m_X ^2)^2}\sqrt{1-{m_{\tau}^2 \over m_X^2}}
\left[ 1+ {1\over 3} {\overrightarrow{p}^2 \over m_{\tau}^2} \right]
\nonumber\\
\sigma_{XX\rightarrow \tilde f_i {\tilde f}_i^*} v &=&
{(\lambda'^2_{Li} +\lambda'^2_{Ri})^2 \over 16 \pi }
\sqrt{1-{m_{\tilde f_i}^2 \over m_X^2}}
{ m_X^2 -m_{\tilde f_i}^2
\over (m_{Y^{lep.}}^2 +m_X^2 -m_{\tilde f_i}^2)^2}\,,\label{annihilation}
\eea
and the corresponding diagrams are shown in Fig.~2. We set $\lambda'_{\tau}\equiv\lambda'_{L\tau}=\lambda'_{R\tau}$
and $\lambda'_i\equiv\lambda'_{L i}=\lambda'_{R i}$.

DM annihilation to squarks requires the squark masses to be
lighter than the DM mass, which in turn is lighter than the connector, which is
bounded by precision electroweak data and perturbativity to be lighter than 600 GeV.
Generally though, in GMSB-inspired models the squarks are heavier than the sleptons
due to the hierarchy of the gauge couplings. Thus, such light squark masses are not natural.
Besides, the cascade products of heavy squarks and the resulting neutrino spectra are
extremely model dependent, being determined by the features of the sparticle spectrum. For these reasons we decide to
avoid the treatment of annihilation to squarks and focus on the channels dominated by leptonic
Yukawa couplings.

\section{Annihilation Products and Propagation}

\subsection{DM Annihilation in the Sun}

The time dependence of the number of DM particles in the Sun is determined by the balance of capture
and annihilation. A massive particle can be gravitationally captured by the Sun from the galactic halo at a
rate $C_{\odot}$.
On the other hand, annihilation at the center of the Sun will decrease the number of particles at a rate $\Gamma_{A}$.
Thus, the time evolution of the number of DM particles in the Sun $N$ is given by the solution to the
differential equation~\cite{Jungman:1995df},
\begin{equation}
 \frac{dN}{dt}=C_{\odot}-C_{A}N^{2}\,,\label{Nrate}
\end{equation}
where $C_{A}=2\Gamma_{A}/N^2$.

Assuming that the number of collisions DM particles undergo
inside the Sun during the Sun's lifetime is large enough for them to thermalize~\cite{Griest:1986yu}, 
the number density of DM particles at a
distance $r$ from the solar core can be expressed in terms of the DM mass $m_{X}$,
the temperature of the Sun $T$ and the gravitational potential $\phi(r)=2\pi\rho r^2/(3M_{Pl}^2)$:
\begin{equation}
 n(r)=n_{0}e^{-\phi(r)\frac{m_{X}}{T}}=n_{0}e^{-\frac{2\pi\rho r^2}{3M_{Pl}^2}\frac{m_{X}}{T}}\,,\label{DMdensity}
\end{equation}
where $n_{0}$ is the DM number density at the center of the Sun,
$\rho$ is the average density of the Sun and $M_{Pl}=1.22\times10^{19}$~GeV is the Planck mass.
$C_{A}$ is then a constant that depends only on $\rho$, $T$ and the
annihilation cross section $\langle\sigma_A v\rangle$ averaged over the velocity distribution
in the limit $v\rightarrow 0$:
\begin{equation}
 C_{A}=\frac{\int n(r)^2 \langle\sigma_A v\rangle d^3r}{(\int n(r) d^3r)^2}=
 \langle\sigma_A v\rangle\left(\frac{3M_{Pl}^2 T}{m_{X}\rho}\right)^{-\frac{3}{2}}\,.\label{Ca}
\end{equation}
Equation~(\ref{Nrate}) admits the solution,
\begin{equation}
 N(t)=\sqrt{\frac{C_{\odot}}{C_{A}}}\tanh\left(\frac{t}{\tau}\right)\,,\label{Nsolution}
\end{equation}
where $\tau\equiv1/\sqrt{C_{A}C_{\odot}}$ is the characteristic time necessary to reach equilibrium.
For $t \sim 4.5\,{\rm Gyr}$ and  $\langle \sigma_{ann.} v \rangle \sim 1\,{\rm pb}$ (as is the case for WIMPless dark matter) 
and $\sigma_{SD} \gsim 10^{-6}\,{\rm pb}$ (the range of interest for IceCube/DeepCore), it is 
known~\cite{Jungman:1995df,Griest:1986yu} that $t / \tau \gg 1$ for the range \mbox{$150\,{\rm GeV} < m_X < 600\,{\rm GeV}$} 
considered here.  
Since $\Gamma_{A}=C_{A}N^2/2=C_{\odot}\tanh^{2}(t/\tau)/2$, one can easily see that when $t\gg\tau$,
$\Gamma_{A}\rightarrow
\Gamma_{eq}\equiv C_{\odot}/2$.

This limit simplifies the calculations,
as it relates the event rate to the DM-nucleon scattering cross section,
thus bypassing all the astrophysical uncertainties related to
the solar model. However, the condition of equilibrium is not guaranteed.
For example, in Ref.~\cite{Ellis:2009ka} it was recently shown that there are regions of mSUGRA parameter
space for which the
annihilation rate is far below the capture rate. Moreover, since DM can annihilate to hidden
sector particles as
well as MSSM particles, the annihilation rate relevant for neutrino detection is scaled by the branching
fraction to
MSSM decay products, $\Gamma_A^{MSSM} = \Gamma_A B_F^{MSSM}$.
In the figures presented in Sec.~4, we account for these uncertainties by scaling the muon event rates at
the detector by the parameter \mbox{$\xi\equiv\Gamma_{A}^{MSSM}/\Gamma_{eq}  = B_F^{MSSM} \tanh^2(t/\tau)$.}
However, our discussion assumes $\xi=1$, as this choice
allows us to draw quantitative conclusions. Event rates
at conditions away from equilibrium can be obtained by simply choosing a lower value for $\xi$.

\subsection{Capture Rate}

If WIMPless DM is Majorana, it has only SD scattering.
This property presents the advantage of not being tied to the direct-detection bounds on the SI
cross section. The corresponding capture rate is~\cite{Jungman:1995df,Gould:1987ir},
\begin{equation}
 C_{\odot}\simeq3.35\times10^{20}\textrm{ s}^{-1}\left(\frac{\sigma_{SD}}{10^{-6}\textrm{ pb}}\right)
 \left(\frac{100\textrm{ GeV}}{m_{X}}\right)^{2}\,,\label{capture}
\end{equation}
where we have taken the local DM density to be 0.3~GeV/cm$^3$, and the root-mean-square of the velocity dispersion
in the halo to be 270~km/s. For SD scattering in the Sun, the only relevant nucleus with spin is hydrogen.

\subsection{Neutrino Spectra}

We limit our consideration to the cases of DM particles annihilating to taus, staus and sneutrinos of
all three families.  For the cases where DM annihilates to sparticles, we assume that the
annihilation product is the next-to-next-to-lightest SUSY particle (NNLSP),
and the lightest neutralino is the next-to-lightest SUSY particle (NLSP);
in GMSB the lightest SUSY particle (LSP) is the gravitino.
Such a choice is consistent with generic sparticle spectra originating
from GMSB.

In Figs.~3a, 3b and 3c, we show the neutrino energy spectra $dN/dE_{\nu}$ for
an annihilation of WIMPless particles of mass $m_{X}=150$~GeV into taus, $X\bar{X}\rightarrow\tau^-\tau^+$. 
The corresponding
antineutrino spectra are shown in Figs.~3d, 3e and 3f. Figures~3a and 3d show the spectra at the origin;
Figs.~3b and 3e at the surface of the Sun and Figs.~3c and 3f at the detector. Similar spectra are obtained for 
annihilation into
taus for the other two benchmark DM masses we consider in this paper: $m_{X}=300$~GeV and $m_{X}=400$~GeV. 
In Fig.~4 we show the
neutrino and antineutrino spectra from annihilation into staus
($X\bar{X}\rightarrow\tilde{\tau}_{1}^-\tilde{\tau}_{1}^+$) of a $m_{X}=300$~GeV DM particle. In Fig.~5, the 
neutrino and
antineutrino spectra from $m_{X}=400$~GeV DM~annihilation into $\tilde{\nu}_{eL}\tilde{\nu}_{eL}$,
$\tilde{\nu}_{\mu L}\tilde{\nu}_{\mu L}$ and  $\tilde{\nu}_{\tau L}\tilde{\nu}_{\tau L}$
with equal branching fractions are shown. The spectral shapes for $m_{X}=150$, $400$~GeV DM~annihilation into 
staus are similar to those in
Fig.~4, while those for $m_{X}=150$, $300$~GeV DM~annihilation into sneutrinos are similar to Fig.~5.

\begin{figure}[ht!]
	\mbox{\ \includegraphics[width=55mm, height=35mm]{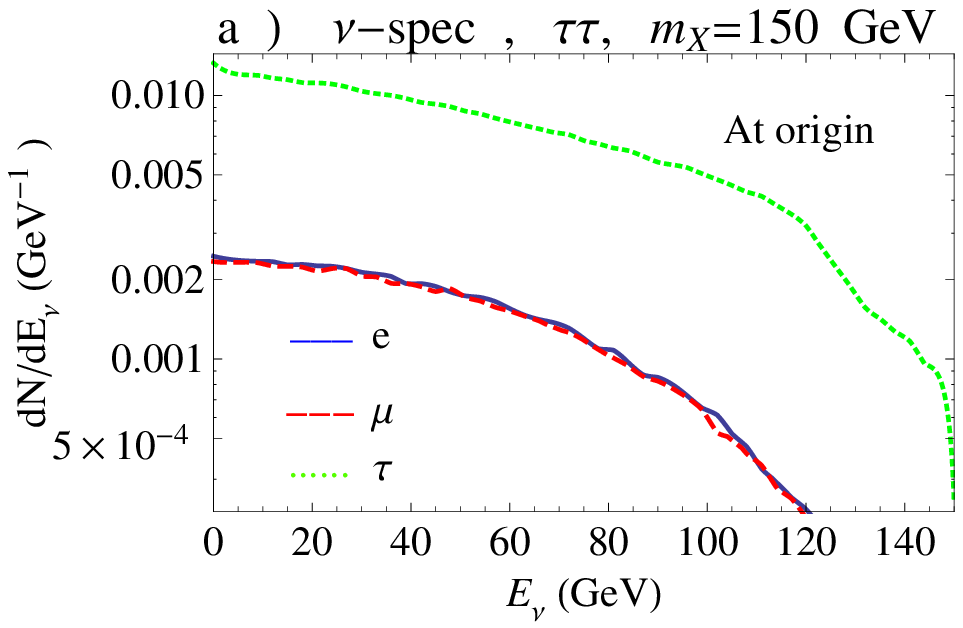}\
\includegraphics[width=55mm, height=35mm]{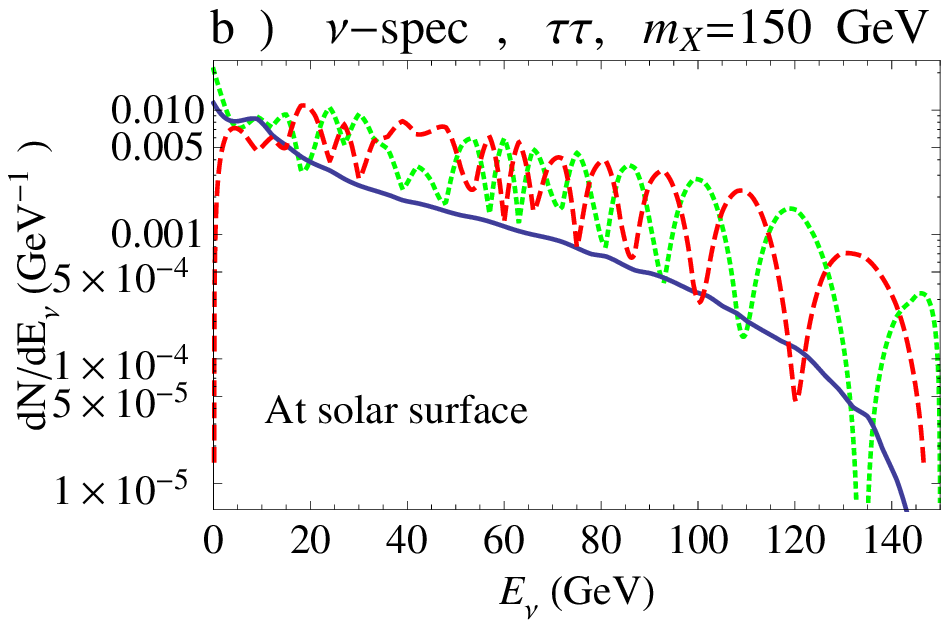}\ \includegraphics[width=50mm, height=35mm]{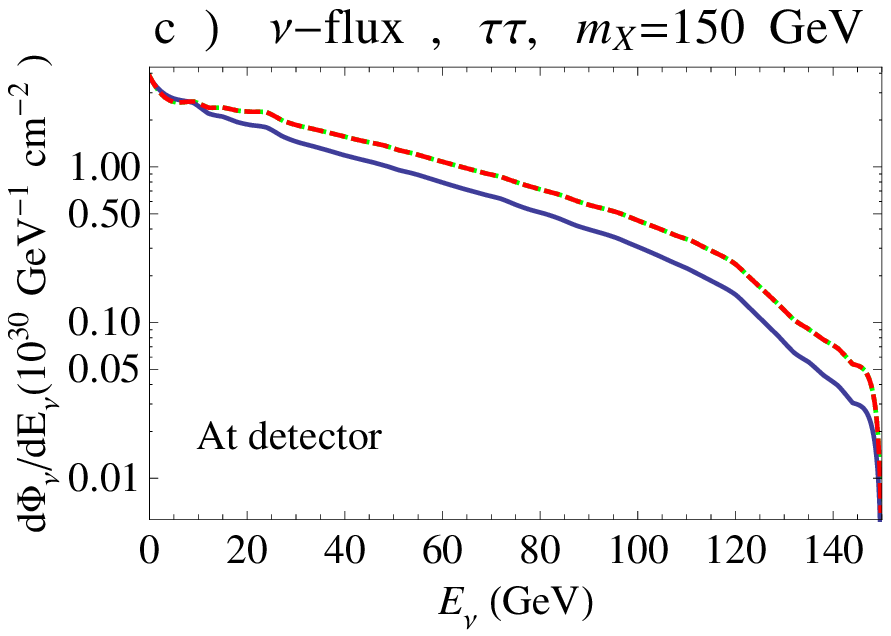}}\\
	\mbox{\ \includegraphics[width=55mm, height=35mm]{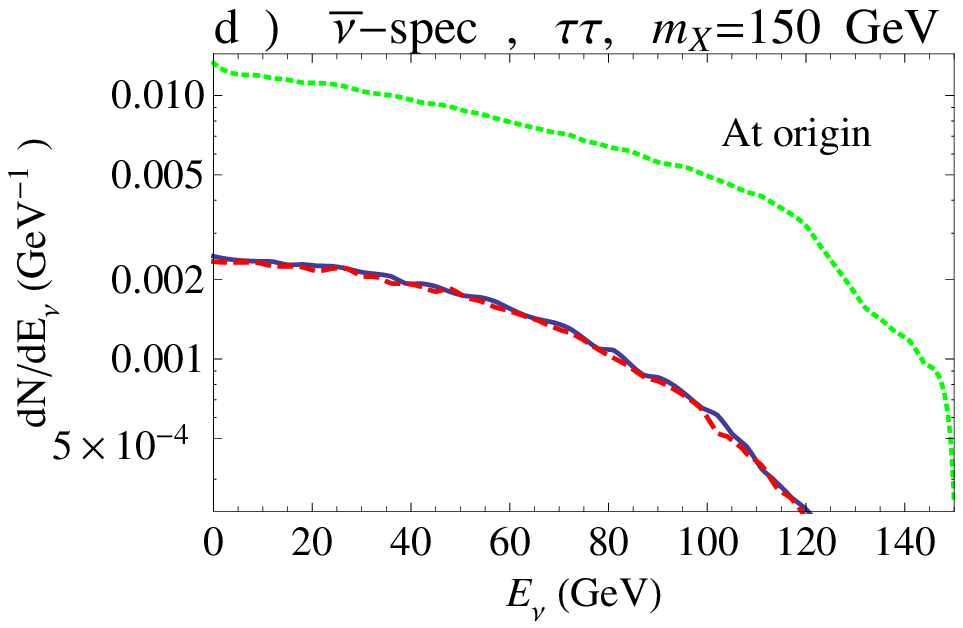}\
\includegraphics[width=55mm, height=35mm]{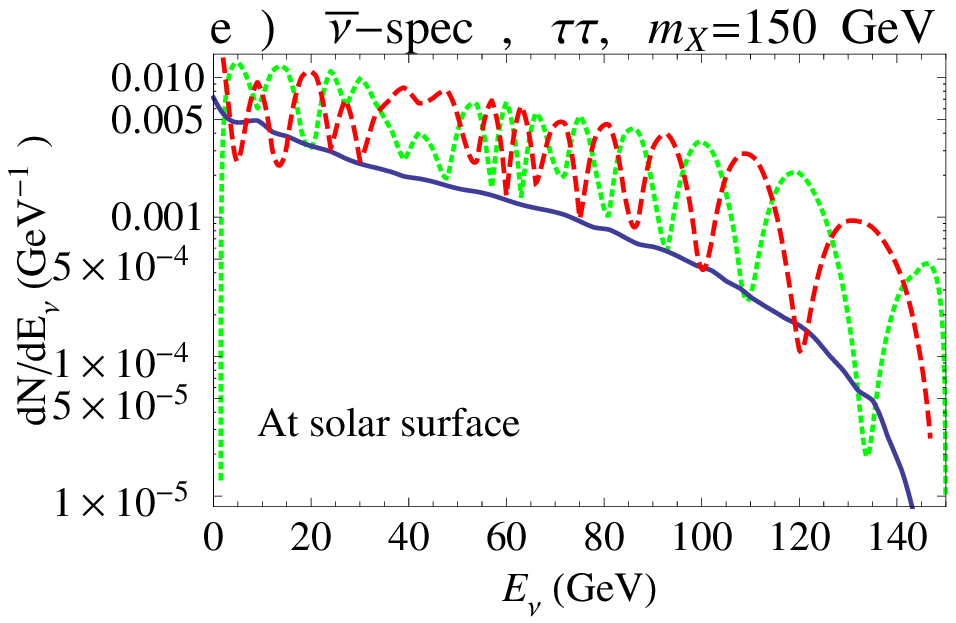}\ \includegraphics[width=50mm, height=35mm]{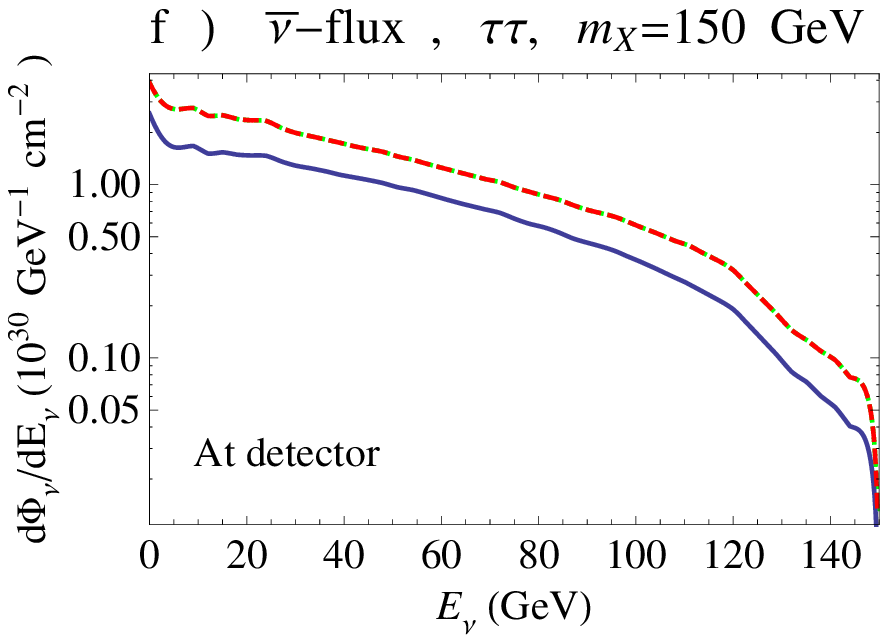}}
	\caption{Neutrino spectra and fluxes for all three flavors per $X\bar{X}\rightarrow\tau^-\tau^+$ 
annihilation with $m_{X}=150$~GeV,
a) at production, b) at
     the surface of the Sun, c) flux $\Phi_{\nu}\equiv N/4\pi R_{SE}^2$ at Earth. d), e), f)
     $\bar \nu$ spectra and fluxes.}
\end{figure}

\begin{figure}[ht!]
	\mbox{\ \includegraphics[width=55mm, height=35mm]{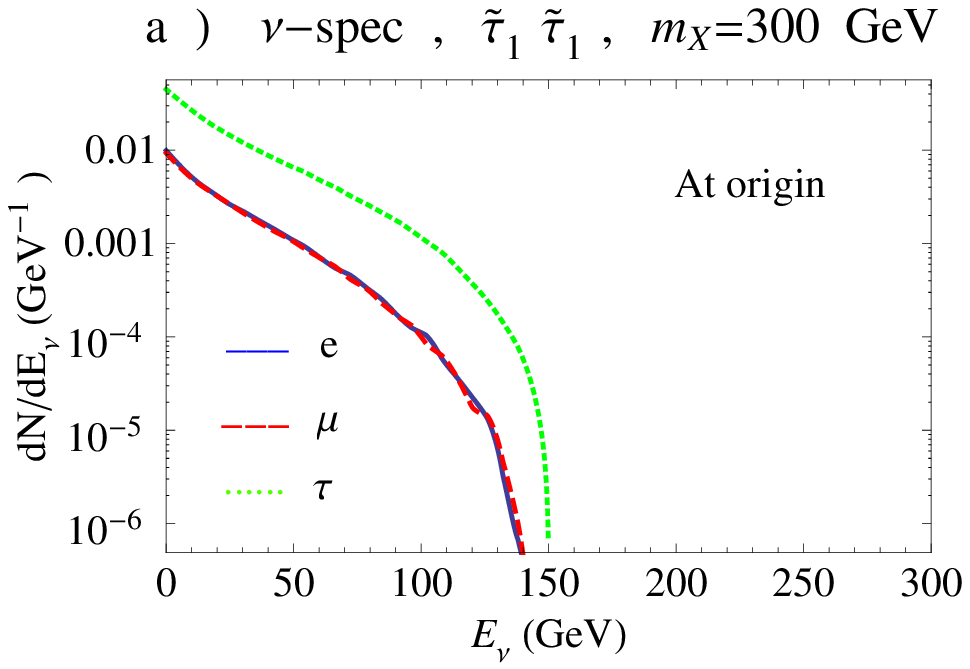}\
\includegraphics[width=55mm, height=35mm]{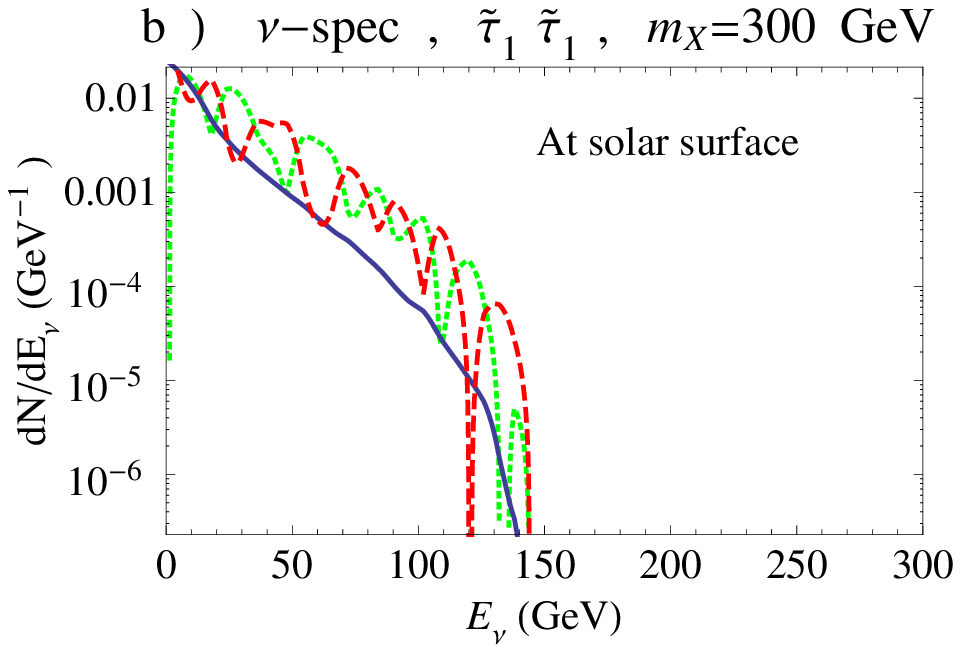}\ \includegraphics[width=51mm, height=35mm]{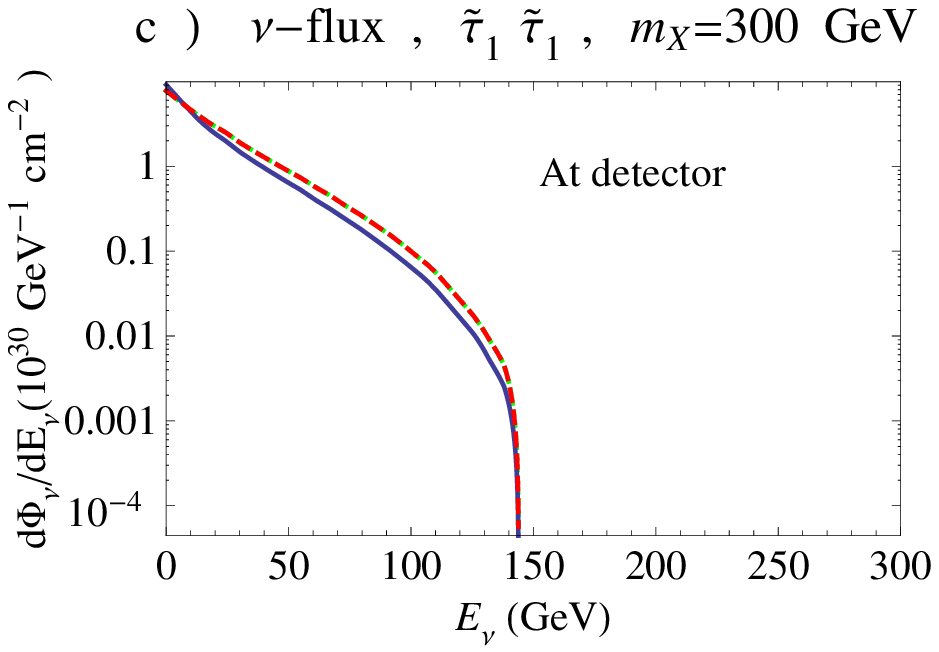}}\\
	\mbox{\ \includegraphics[width=55mm, height=35mm]{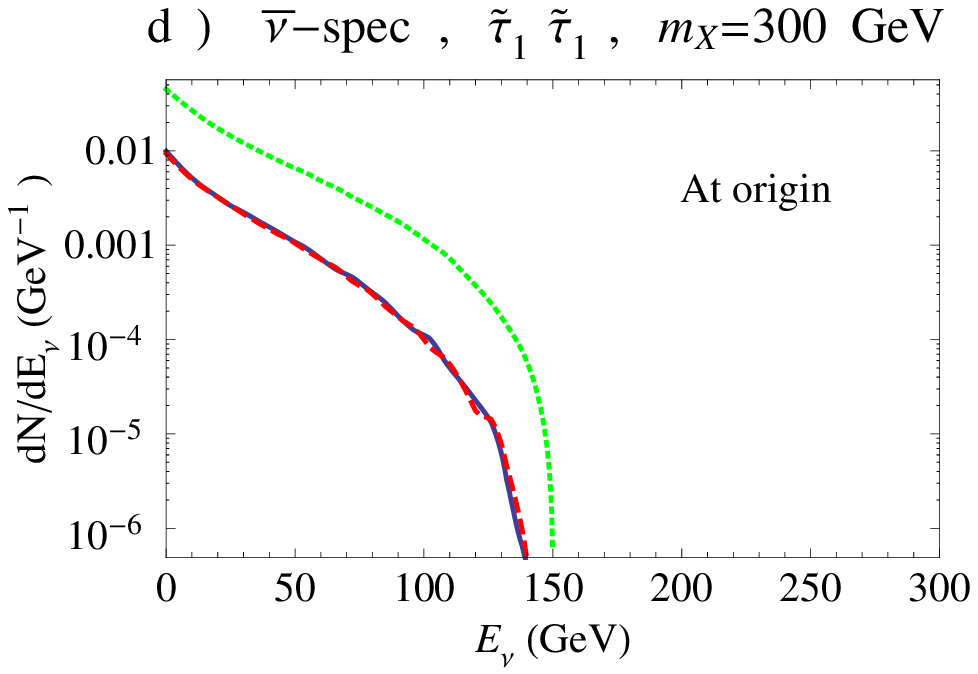}\
\includegraphics[width=55mm, height=35mm]{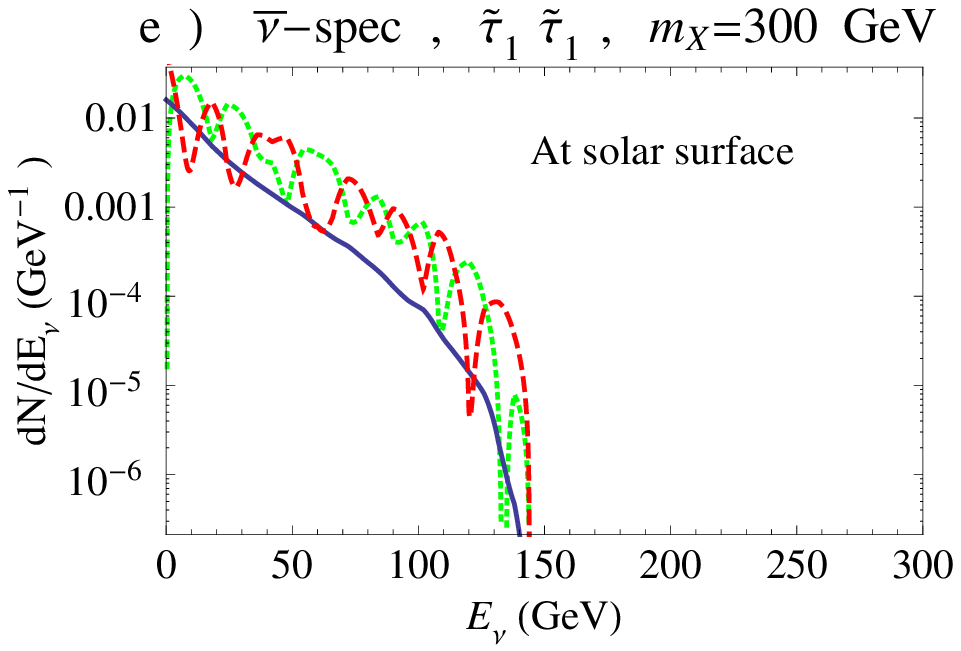}\ \includegraphics[width=51mm, height=35mm]{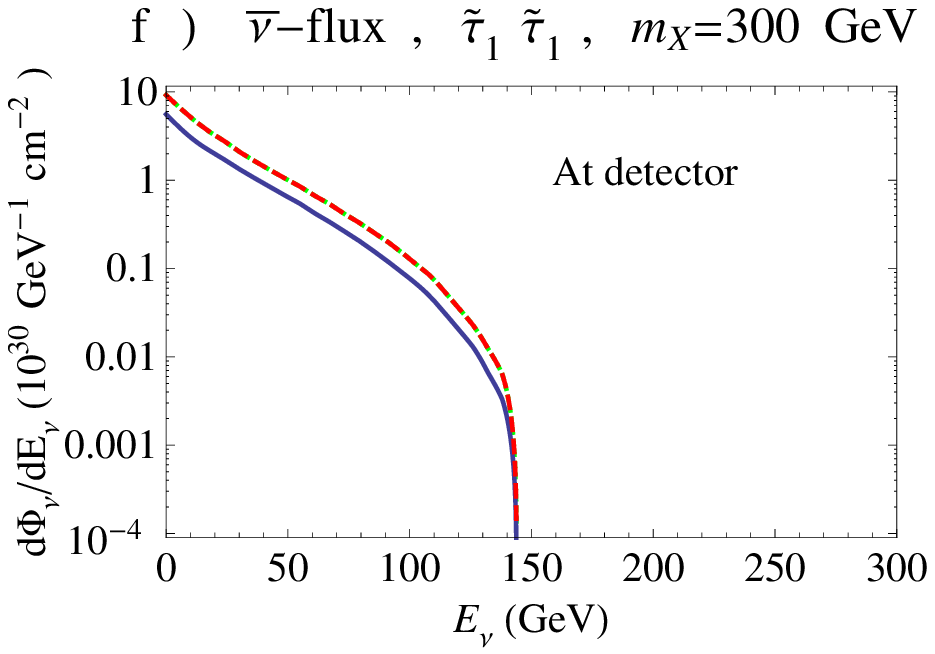}}
	\caption{Similar to Fig.~3 for $X\bar{X}\rightarrow\tilde{\tau}_1^-\tilde{\tau}_1^+$ with $m_{X}=300$~GeV.}
\end{figure}

\begin{figure}[ht!]
	\mbox{\ \includegraphics[width=55mm, height=35mm]{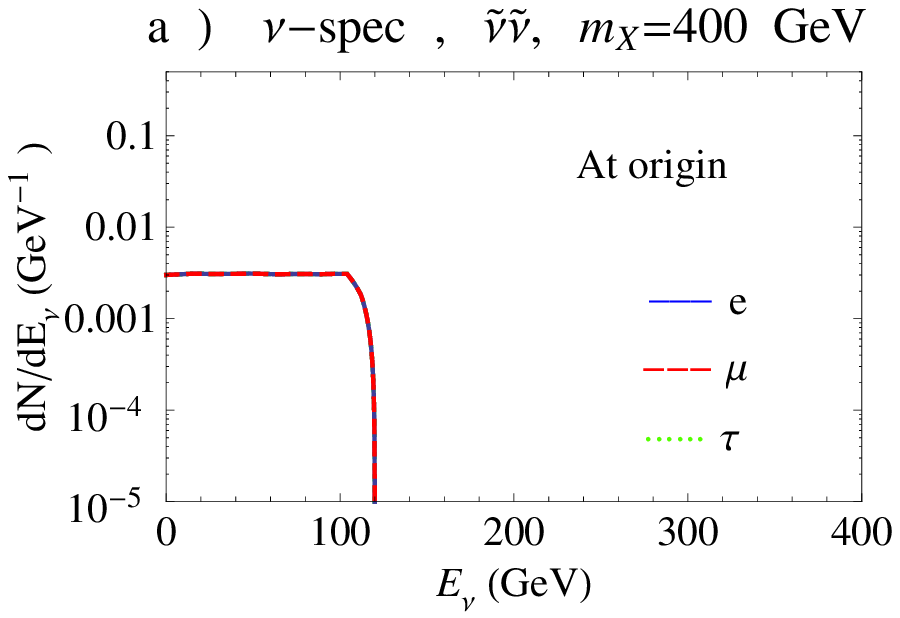}\
\includegraphics[width=55mm, height=35mm]{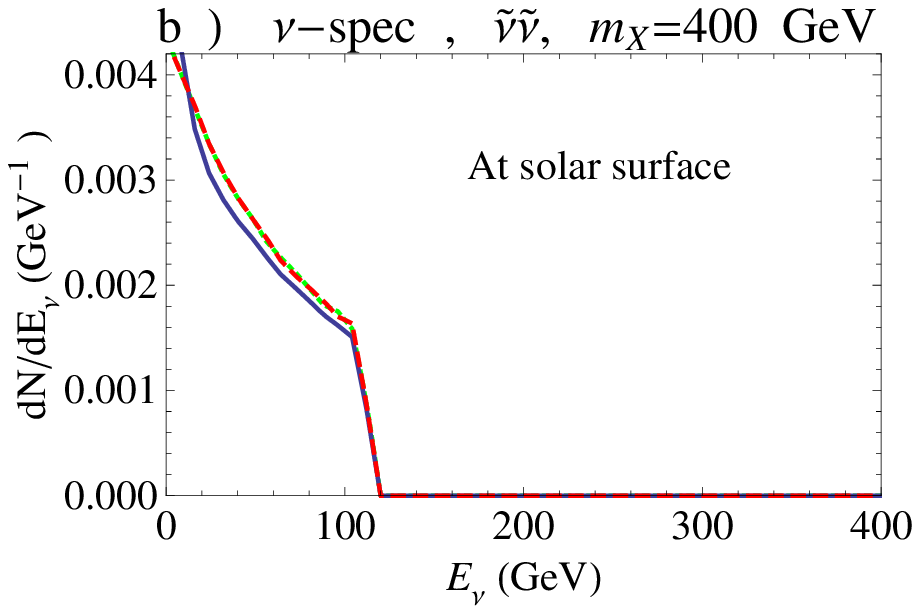}\ \includegraphics[width=50mm, height=35mm]{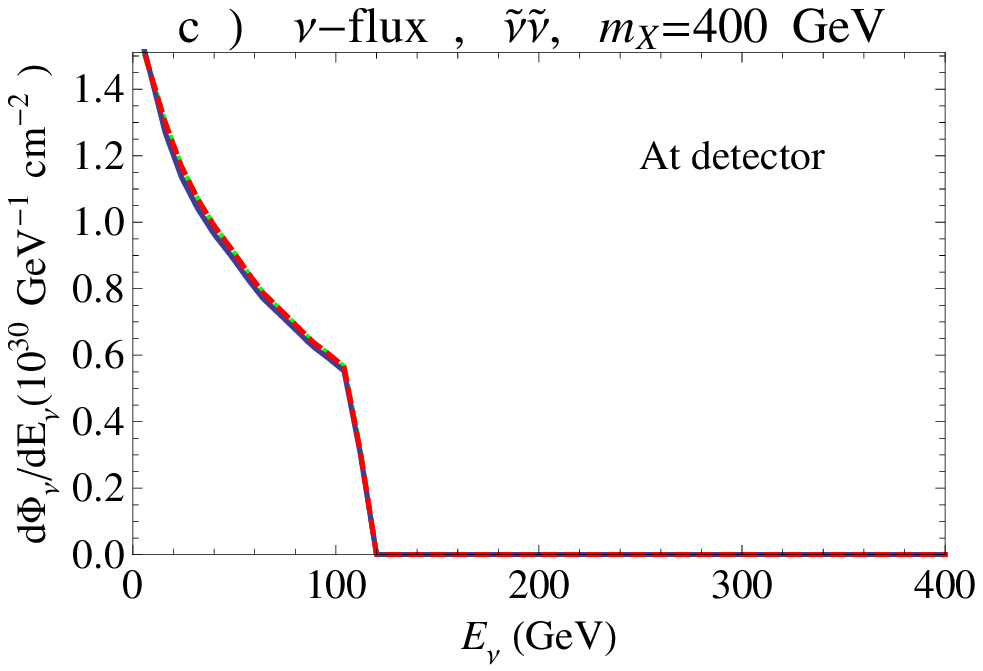}}\\
	\mbox{\ \includegraphics[width=55mm, height=35mm]{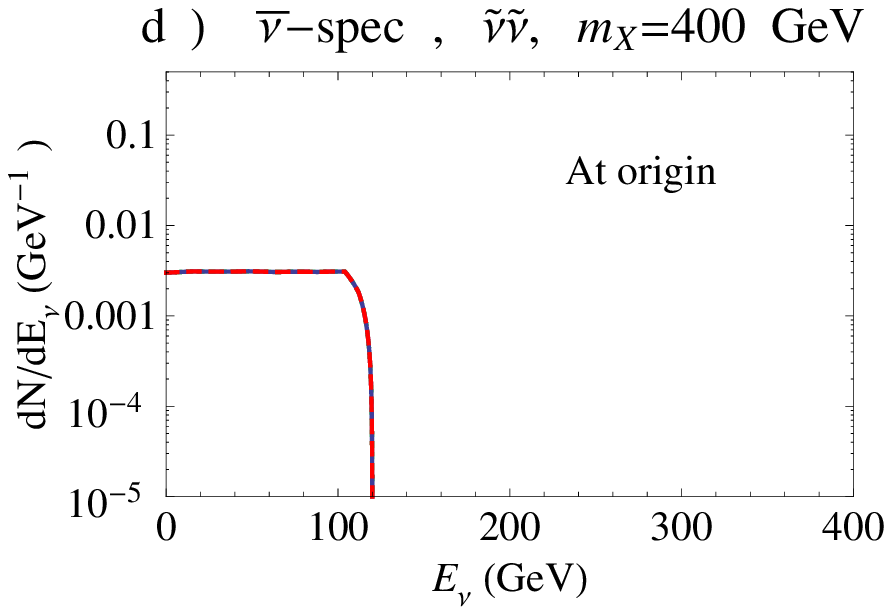}\
\includegraphics[width=55mm, height=35mm]{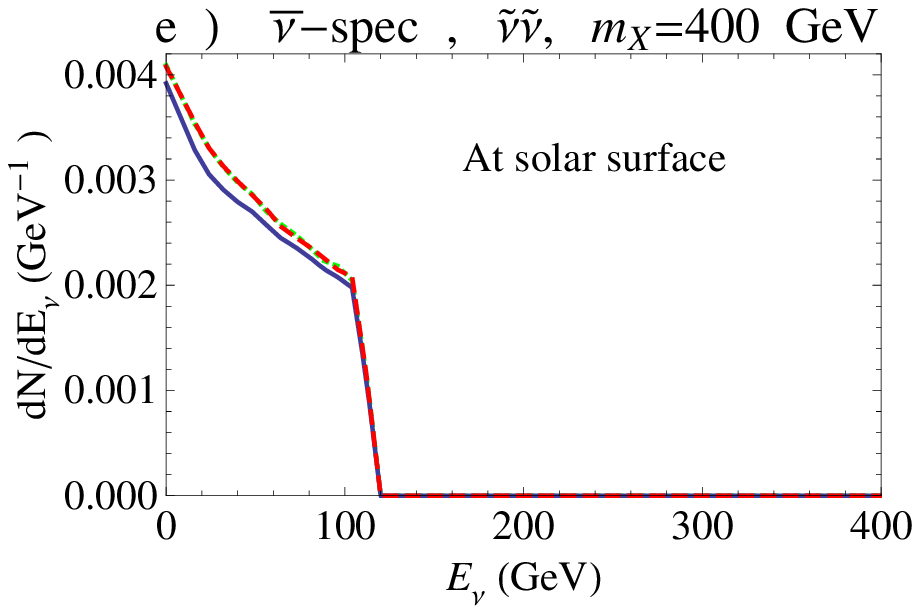}\ \includegraphics[width=50mm, height=35mm]{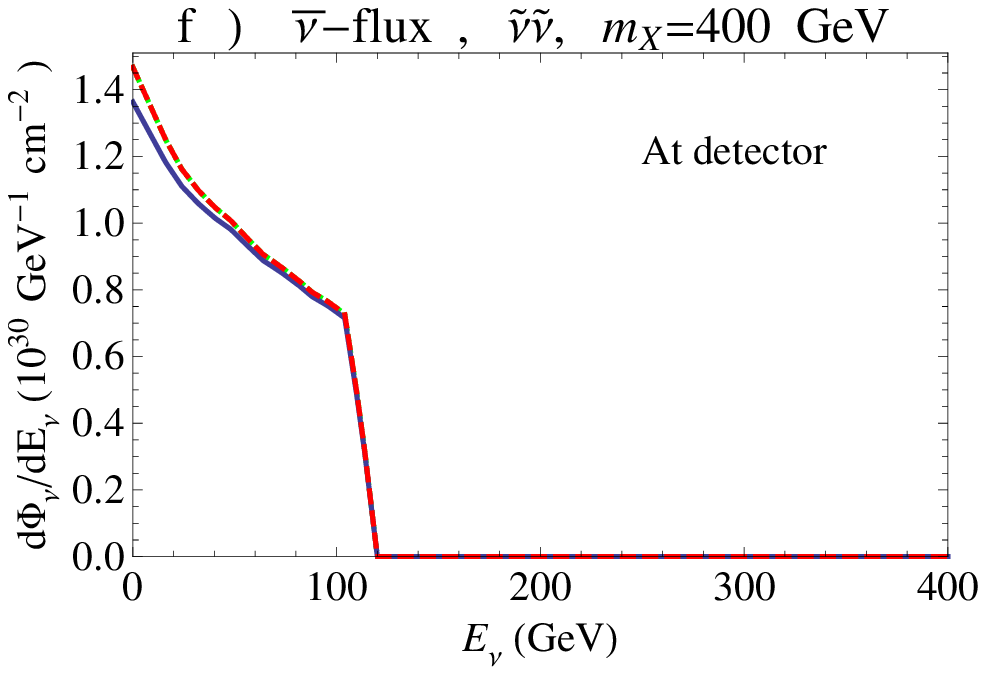}}
	\caption{Similar to Figs.~3, 4 for $X\bar{X}\rightarrow\tilde{\nu}_{e(\mu,\tau)}\tilde{\nu}_{e(\mu,\tau)}$ with equal
branching fractions and \mbox{$m_{X}=400$~GeV}.}
\end{figure}

We obtain the neutrino spectra by cascading the annihilation products
with \mbox{PYTHIA 6.4}~\cite{Sjostrand:2006za}. In the
cases of Figs.~4 and 5 we consider a typical GMSB spectrum, like the one given in Ref.~\cite{Dimopoulos:1996yq}.
We assume \mbox{$m_{\tilde{\tau}_{1}}=137$~GeV} and \mbox{$m_{\tilde{Z}_{1}^{0}}=94.5$}~GeV,
corresponding to \mbox{$\tan\beta\sim10$}. The cascade proceeds primarily through
$\tilde{\tau}_1^{-(+)}\rightarrow\tilde{Z}_1^0\tau^{-(+)}$,
hence the higher number of $\nu_\tau$s and $\bar \nu_\tau$s at the origin in Figs.~4a and 4d.

In the case where the DM candidate
annihilates primarily into sneutrinos (Fig.~5), we assumed for simplicity that the three sneutrino
masses are degenerate ($m_{\tilde{\nu}}=111.5$~GeV) and that the couplings to the three flavors are
identical, so that the production spectra for  the three neutrino families are the same,
$\tilde{\nu}_{eL(\mu L,\tau L)}\rightarrow\tilde{Z}_1^0\nu_{e(\mu,\tau)}$. As in the previous case we have
$m_{\tilde{Z}_{1}^{0}}=94.5$~GeV, corresponding to $\tan\beta\sim10$. Note that since WIMPless DM
particles only annihilate leptonically into SM particles or the NNLSP, the neutrino spectra have the same shape whether $X$
is heavier or lighter than the top quark. Notice that the spectra
from tau decays are broader than in the stau and sneutrino cases. This is expected because of the influence of the neutralino
mass on kinematics. As a matter of fact, the steep drop in the neutrino spectra arising from annihilation to sparticles
will have important consequences for
detection at IceCube and DeepCore, as we discuss in Sec.~4.

The propagation of neutrinos produced at the center of the Sun through the solar medium and to the Earth is
detailed in the appendices.
We take into account neutrino oscillations and the effects of neutral-current (NC) and charged-current (CC)
interactions and tau regeneration.
In Figs.~3b, 3e and corresponding panels of Figs.~4 and 5 we show the neutrino and antineutrino fluxes after
propagation to the surface of
the Sun. They present expected features; for example, an accumulation of events at lower energies is visible, due
to energy losses of the neutrinos undergoing NC and CC~scattering inside the Sun. Tau~regeneration, too, has
the effect of increasing the low-energy neutrino flux. Furthermore, we see that flavor mixing affects the
$\nu_\mu$ and $\nu_\tau$ spectra but leaves the $\nu_e$ spectra unaltered, since $\theta_{23}=45^{\circ}$,
$\theta_{13}=0^{\circ}$
(Eq.~\ref{Hamiltonian} of Appendix~A). Figures~3c, 3f and corresponding panels of Figs.~4 and 5 show the fluxes at the Earth after our
averaging procedure, described in Appendix~A. We define the flux $\Phi_{\nu}\equiv N/(4\pi R_{SE}^2)$, with
$R_{SE}\sim1$~AU being the Sun-Earth distance. Our results are in very good agreement with those of Ref.~\cite{Blennow:2007tw}.
Notice that almost all modulation is washed out.\medskip


The flux of high energy neutrinos produced by DM annihilation can be detected by
neutrino telescopes like IceCube~\cite{Resconi:2008fe}.  The IceCube detector at the
South Pole is an array of 80 strings uniformly spaced 125~m from one another, each with
60 digital optical modules. It is deployed at a depth between 1450~m and 2450~m, and the instrumented
volume of ice covers 1~km$^3$. When muon neutrinos undergo CC~scattering on nucleons in ice
they produce a muon that propagates through the ice. These muons
leave a trail of Cherenkov radiation that can be detected by the optical modules. We are therefore
interested in determining the muon and antimuon flux in and near the detector, taking into account
possible energy losses due to the propagation in ice. DeepCore~\cite{Schulz:2009zz},
an extension of IceCube, is a denser (72~m spacing) array of six strings surrounding one of the central strings
of IceCube, most of them placed below the ``dust layer'' at a depth of 2100~m.
As detailed in Appendix~B, DeepCore's denser strings lower the
neutrino energy threshold with respect to IceCube and thus allow detection of neutrinos from annihilation of
DM particles of a lower mass. Besides, DeepCore is designed to use the outer instrumented volume of IceCube as a veto
on atmospheric muons from above the horizon to a level of one part in $10^6$, thus drastically reducing the background.
Such properties make it a perfect instrument to test the WIMPless scenario in the case of Majorana DM.

DeepCore's muon effective volume becomes insignificant below muon energies of 10~GeV~\cite{DeYoung}. At energies between 10
and 35~GeV DeepCore cannot provide directional information, since only one or two optical
modules will be triggered. Very good directional information is needed to track the Sun throughout the year, so we fix
DeepCore's energy threshold at $E_{min}=35$~GeV~\cite{DeepCoreThreshold}.  For
IceCube we consider the very optimistic threshold $E_{thr}=100$~GeV.

\section{Results}

In this section we investigate the prospects for DM detection at IceCube and DeepCore.  We consider a 3$\sigma$-detection as
\begin{equation}
 ``3\sigma\textrm{-detection''}\leftrightarrow\frac{N_{\mu}}{\sqrt{N_{BG}}}=3\,,\label{signtobg}
\end{equation}
where $N_{\mu}$ is the observed number of muon events and $N_{BG}$ is the number of muon events due to the
atmospheric neutrino background~\cite{Honda:2006qj}, which is discussed further in Appendix C.

For the IceCube detector we divide the muon events
into \textit{upward} and \textit{contained} events, following the treatment of~\cite{Erkoca:2009by}. As we discuss
in Appendix B, the former are due to upward going neutrinos interacting outside the detector volume, the latter
to neutrinos that interact within the instrumented volume. To evaluate the event rate we track the Sun in the sky at different times of the year,
and consider only
events detected when the Sun is below the horizon. The distinction between upward and contained events does not apply to DeepCore, which only
sees contained
events. For DeepCore, IceCube will veto atmospheric
muons from above the horizon to a level of one part in $10^6$. Therefore, at DeepCore we consider the signal detected
throughout the year. We consider a period of 5~yrs of observation. As anticipated in Sec.~3, we rescale the event rates obtained at equilibrium by
$\xi\equiv\Gamma_A^{MSSM}/\Gamma_{eq}$.

\begin{figure}[ht!]
	\mbox{\ \includegraphics[width=53mm, height=40mm]{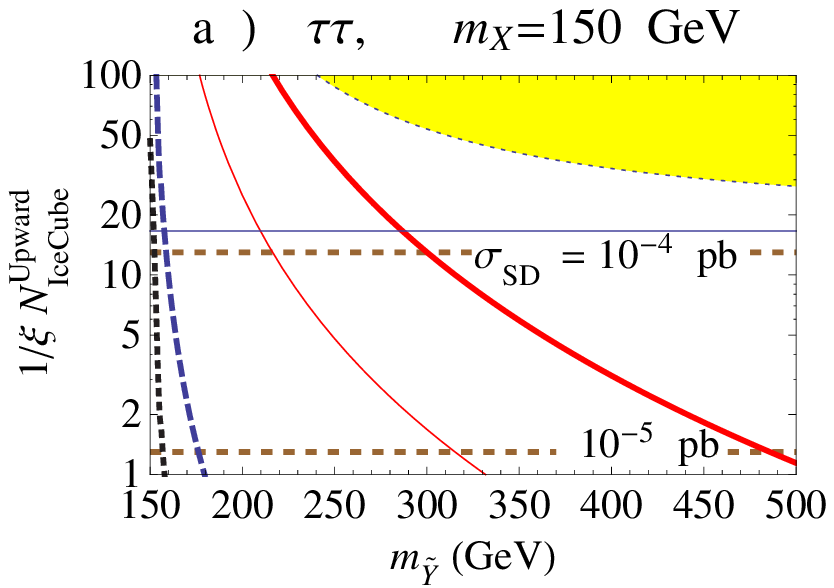}\ \includegraphics[width=53mm, height=40mm]{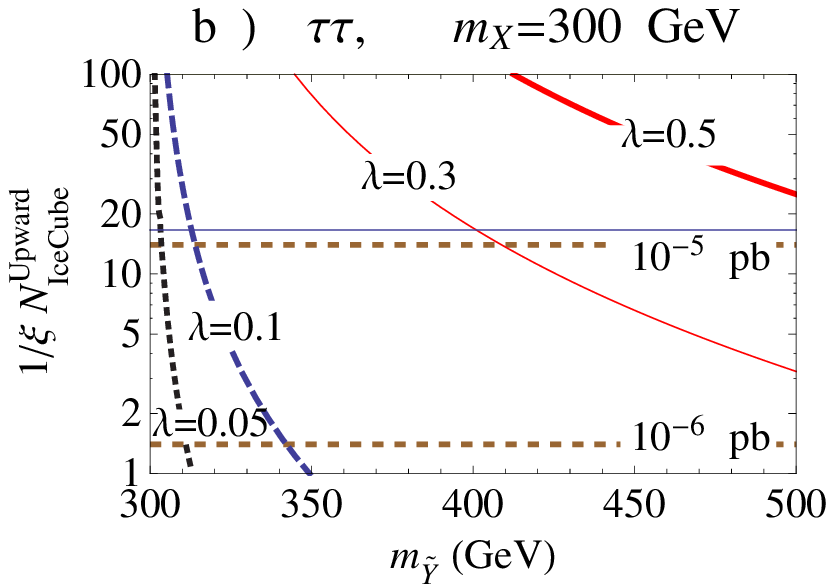}\
\includegraphics[width=53mm, height=40mm]{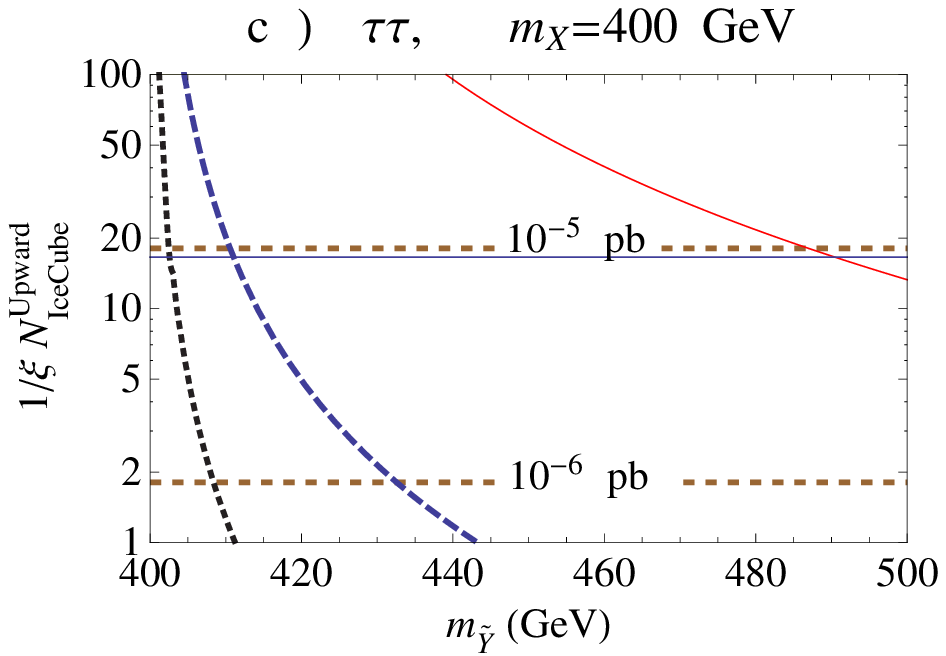}}\\
	\mbox{\ \includegraphics[width=53mm, height=40mm]{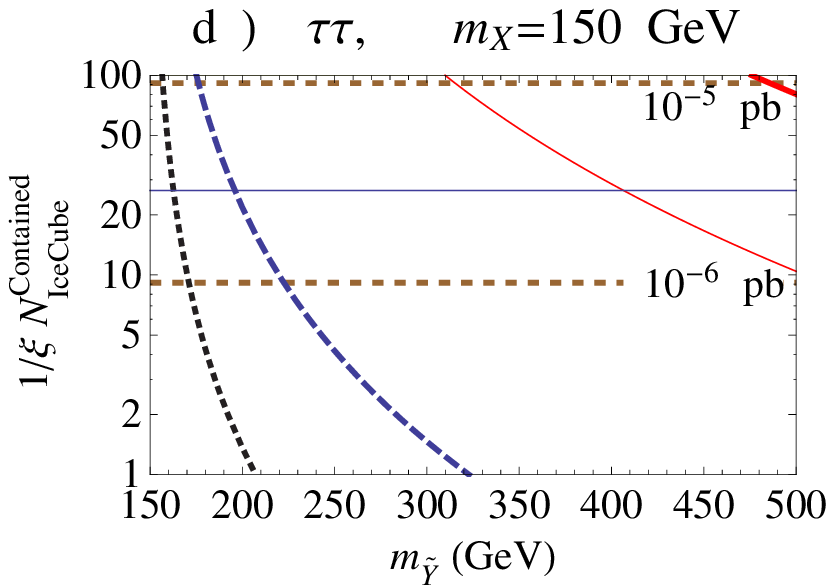}\ \includegraphics[width=53mm, height=40mm]{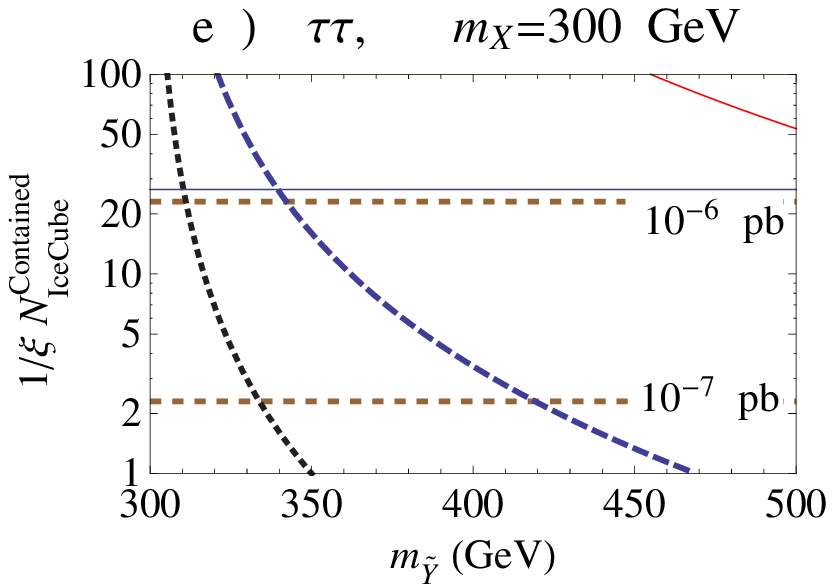}\
\includegraphics[width=53mm, height=40mm]{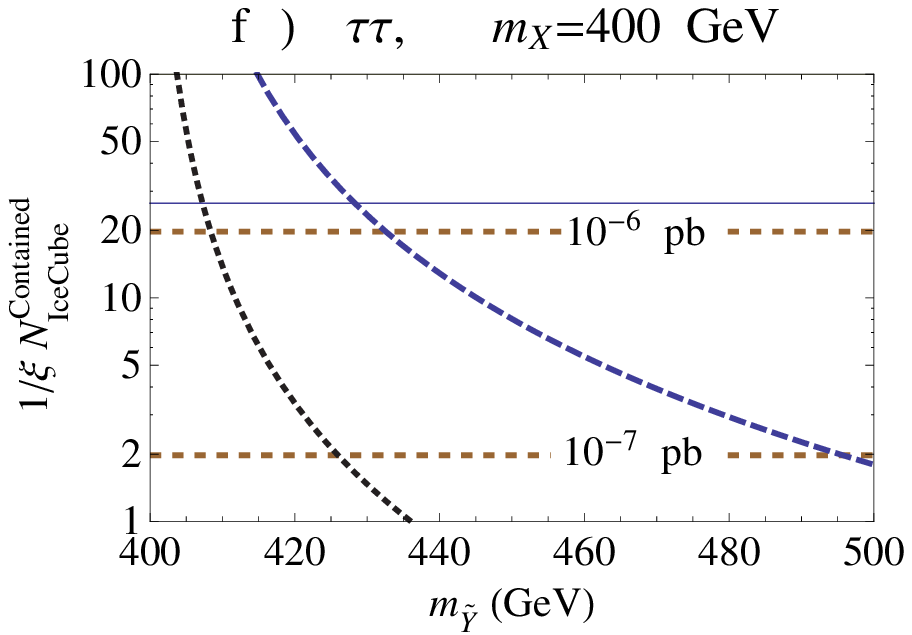}}
	\caption{Number of upward and contained events in 5~yrs of observation at IceCube for three values of $m_{X}$ as a
function of $m_{{\tilde Y}}$
     for the $\tau^-\tau^+$~channel for several choices of $\lambda=\lambda_{(L,R)(u,d)}$.
     Dotted black:~$\lambda=0.05$; dashed blue:~$\lambda=0.1$; solid thin
     red:~$\lambda=0.3$; solid thick red:~$\lambda=0.5$. The blue horizontal lines represent the number of events needed for a
     3$\sigma$~discovery in 5~yrs of observation. The dashed brown horizontal lines
     represent $\sigma_{SD}$ in pb. The yellow band represents values of $\lambda$ that yield the correct relic abundance.}
\end{figure}

\subsection{Event Rates: Tau channel}

In Fig.~6 we plot the upward and contained event rates at IceCube for our three benchmark
DM masses for the $XX\rightarrow\tau^-\tau^+$ channel,
as a function of the scalar connector mass $m_{{\tilde Y}}$ and the Yukawa couplings. We assume that the masses
of the up-type and down-type squarks $\tilde Y$ are degenerate, and that $\lambda\equiv\lambda_{(L,R)(u,d)}$. In
Fig.~7
we plot the contained event rates at DeepCore for the same channel. The event rates [see Eqs.~(\ref{Nevents}),
(\ref{NeventsC})
and (\ref{NeventsDC}) of Appendix~B] are
evaluated for values of the Yukawa couplings which allow calculations in the perturbative regime. The values in pb of
SD cross sections that correspond to the event rates are shown as horizontal dashed brown lines, and labeled on the plots.
Notice the following features:

$\bullet$ As we show in Appendix C, the atmospheric background to upward events in IceCube is
$N_{BG}^{Up}\simeq6.1$~yr$^{-1}$, while the background to contained
events is $N_{BG}^{Con}\simeq15.6$~yr$^{-1}$. The background to contained events at DeepCore is
$N_{BG}^{DC}\simeq2.5$~yr$^{-1}$. The
blue horizontal lines in \mbox{Figs.~6 and 7} indicate the number of signal events needed for a 3$\sigma$~detection in 5~yrs.
We see that a \mbox{$3\sigma$-detection} in upward events at IceCube is possible
for reasonable values of $\lambda$ and $m_{\tilde Y}$. The contained event sample offers even better prospects.

\begin{figure}[ht!]
	\mbox{\ \includegraphics[width=53mm, height=40mm]{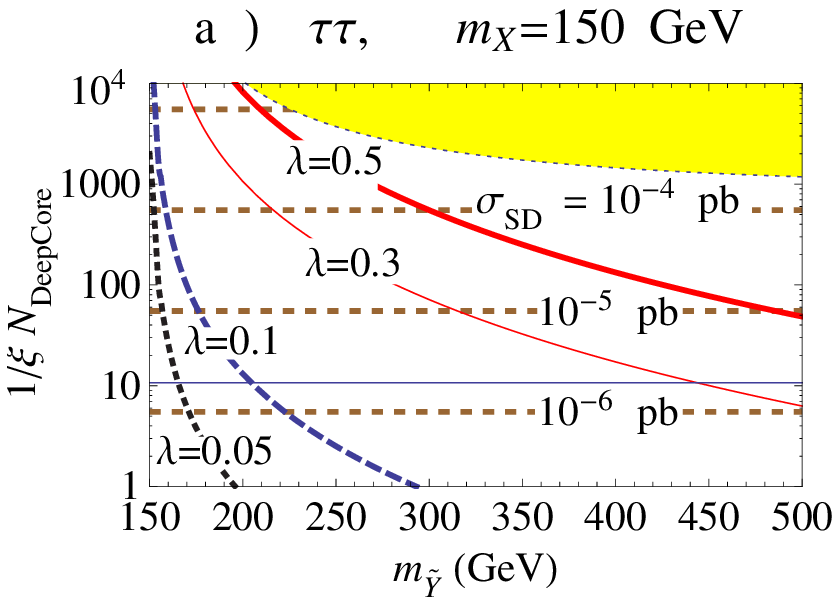}\ \includegraphics[width=53mm, height=40mm]{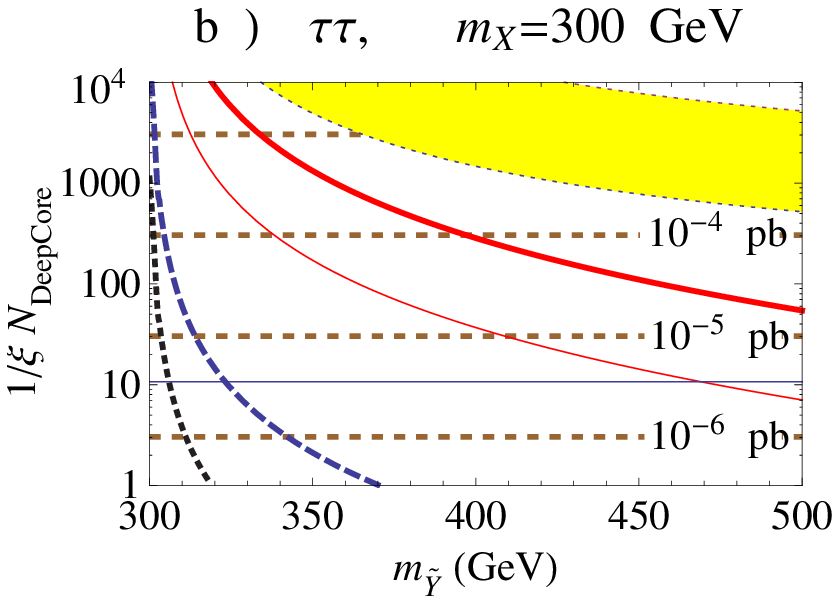}\
\includegraphics[width=53mm, height=40mm]{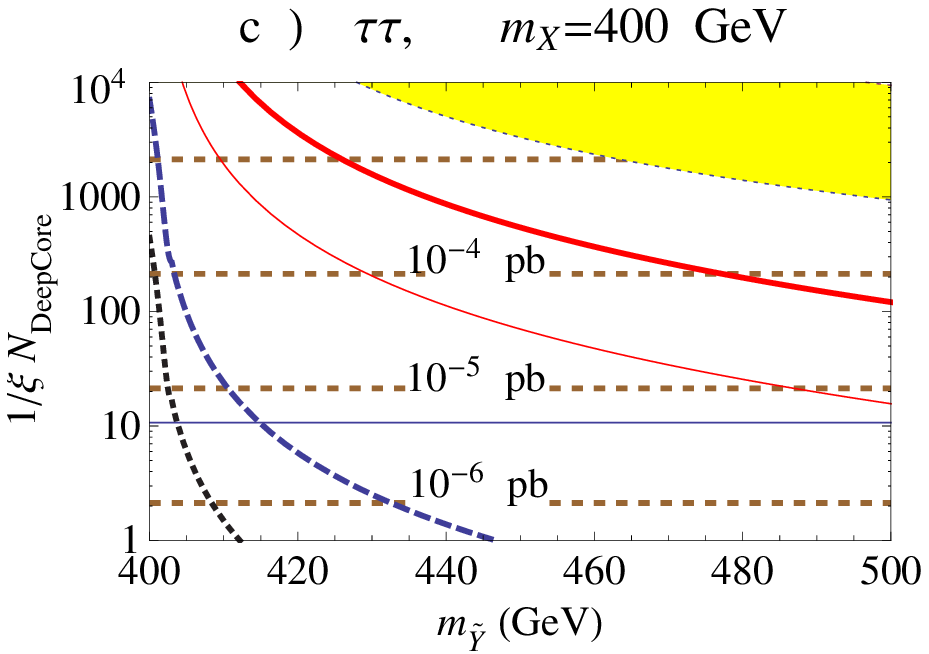}}
	\caption{Similar to Fig.~6 for contained events at DeepCore.}
\end{figure}

$\bullet$ In Fig.~7a we see that DeepCore significantly improves the prospects for observation
of events for $m_{X}=150$~GeV. A 3$\sigma$~detection in this channel can be obtained with Yukawa couplings of
about 0.3 without a need for resonant enhancement. If the Yukawa couplings are allowed to assume larger values,
the prospects for signal detection become even more robust.

$\bullet$ For DM masses well above $E_{thr}$ the advantages of DeepCore
in the $\tau$~channel are less evident. As a matter
of fact a 3$\sigma$~detection is more likely in the contained events at IceCube. For such
high masses the propagator suppression of the SD cross section reduces the flux at the detector and makes the larger
effective volume
of IceCube advantageous.\medskip

One may ask whether it is possible to relate the size of the Yukawa coupling $\lambda$ responsible for SD
scattering to the size of the Yukawa couplings $\lambda_{\tau}$ that appear in Eq.~(\ref{annihilation}), and are responsible
for the annihilation to SM particles. This is possible under certain assumptions:

$\bullet$ The yellow bands in Figs.~6 and 7 indicate the region with $0.1~{\rm pb} \leq \sigma_{X\bar{X}\rightarrow\tau\bar{\tau}}v
\leq 1~{\rm pb}$, assuming
$\lambda'_{\tau}=\lambda$ and $m_{\tilde{Y}^{lep.}}=m_{\tilde{Y}}$. For this range of $\sigma_{X\bar{X}\rightarrow\tau\bar{\tau}}v$
(assuming there are
light bosons in the hidden sector),
annihilation to MSSM particles has a large enough branching
fraction to permit observation at neutrino telescopes, while not being so large as to dilute the relic abundance and
thwart the WIMPless miracle.  Note however, that this constraint is only relevant if DM annihilation
to the hidden sector is not chirality/$p$-wave suppressed.

$\bullet$ If $\lambda'_{\tau}/\lambda >1$,
the yellow band shifts down to lower values. Such a situation may be desirable in the $\tau^-\tau^+$~channel so that the iso-$\lambda$
curves pass through the yellow bands, i.e., the relic abundance and observable signals at neutrino telescopes can be simultaneously obtained
for more of the parameter space.

$\bullet$ As is clear from inspection of Eq.~(\ref{MajSD}) and the first of Eqs.~(\ref{annihilation}), the capture
rate depends on $\lambda$ and $m_{\tilde Y}$ while the annihilation cross section
to taus depends on $\lambda'_\tau$ and $m_{{\tilde Y}^{lep.}}$.
$\xi$ depends on the details of the solar model, particularly on the
ratio $T/\rho$ in Eq.~(\ref{Ca}), and on the
ratios $\lambda'_{\tau}/\lambda$ and $m_{\tilde Y}/m_{\tilde Y^{lep.}}$. A detailed analysis of these aspects
is beyond the scope of this paper.

\begin{figure}[ht!]
	\mbox{\ \includegraphics[width=80mm, height=53mm]{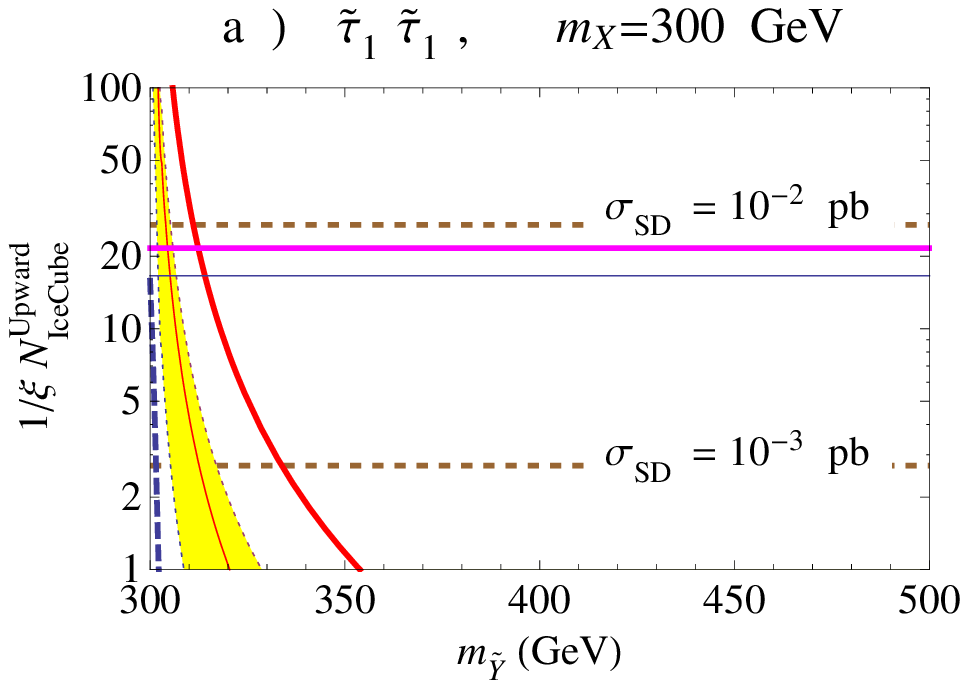}\ \includegraphics[width=80mm, height=53mm]{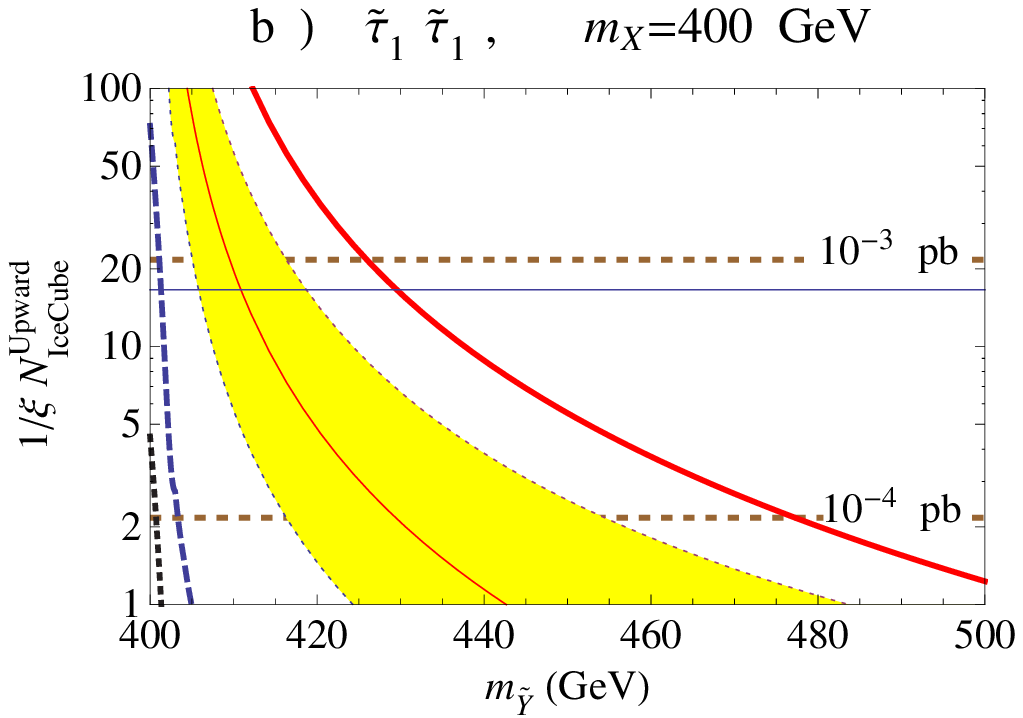}}\\
\mbox{\ \includegraphics[width=80mm, height=53mm]{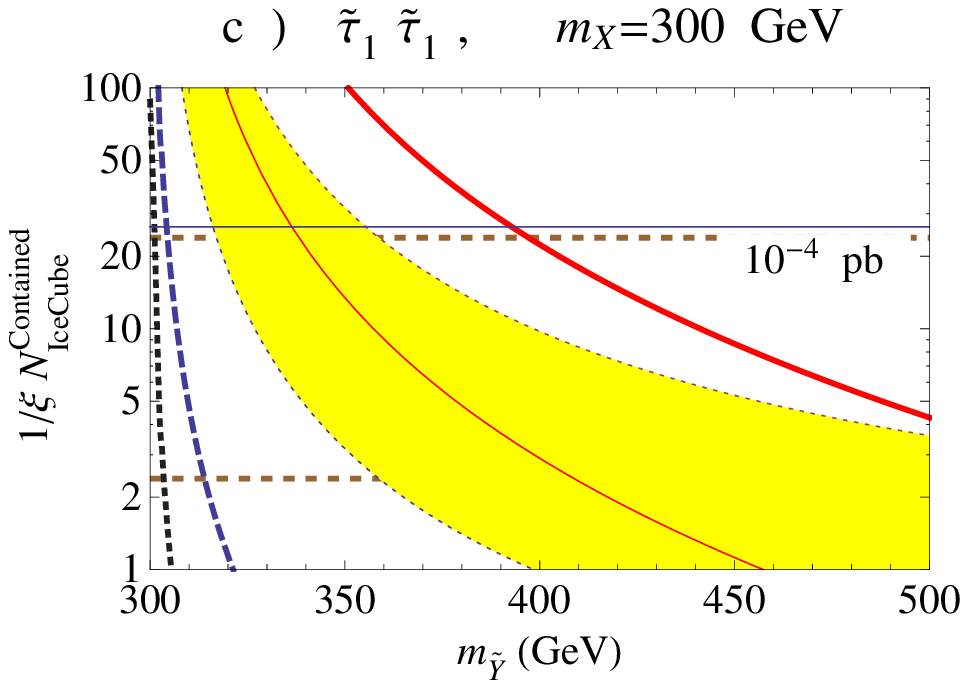}\ \includegraphics[width=80mm, height=53mm]{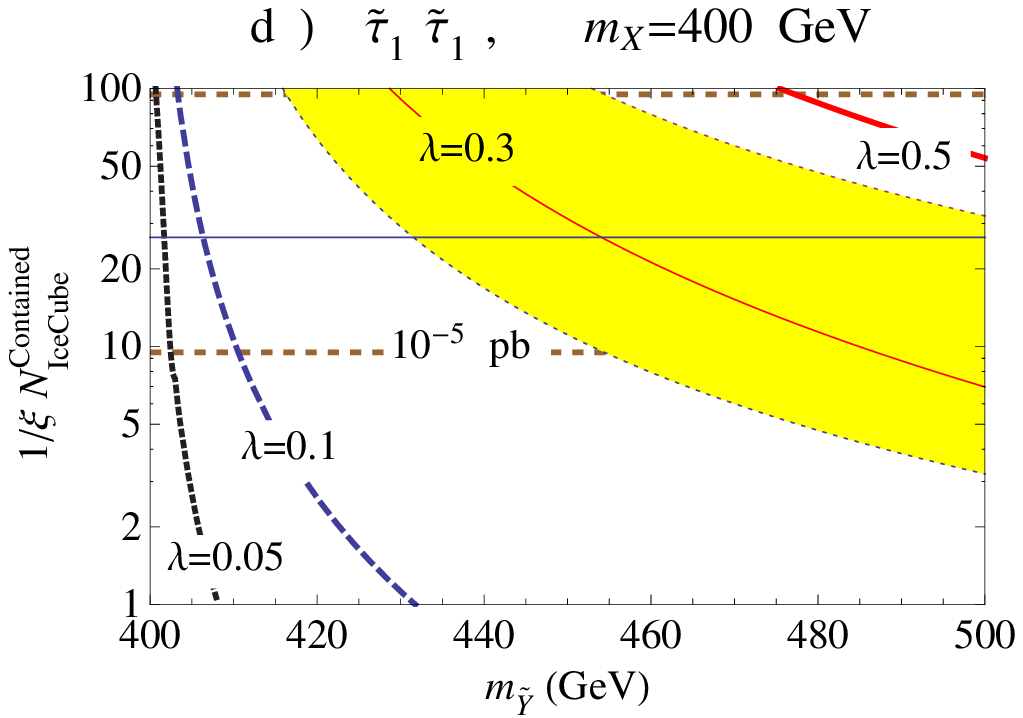}}
	\caption{Similar to Fig.~6 for the $\tilde{\tau}_1^-\tilde{\tau}_1^+$~channel. IceCube is insensitive to this channel for
\mbox{$m_{X}=150$~GeV}. The thick pink horizontal line represents the Super-K 90\% C.~L. upper bound on $\sigma_{SD}$.}
\end{figure}

\begin{figure}[ht!]
	\mbox{\ \includegraphics[width=53mm, height=40mm]{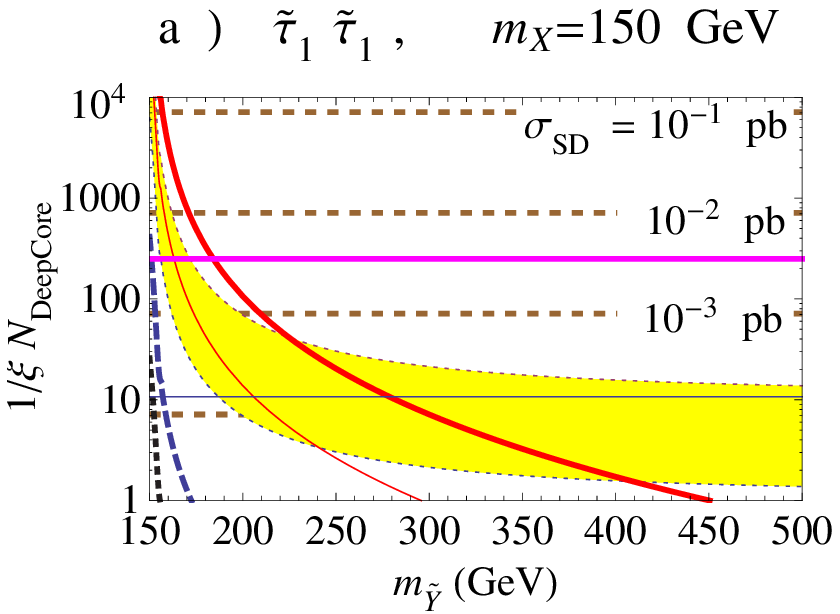}\ \includegraphics[width=53mm, height=40mm]{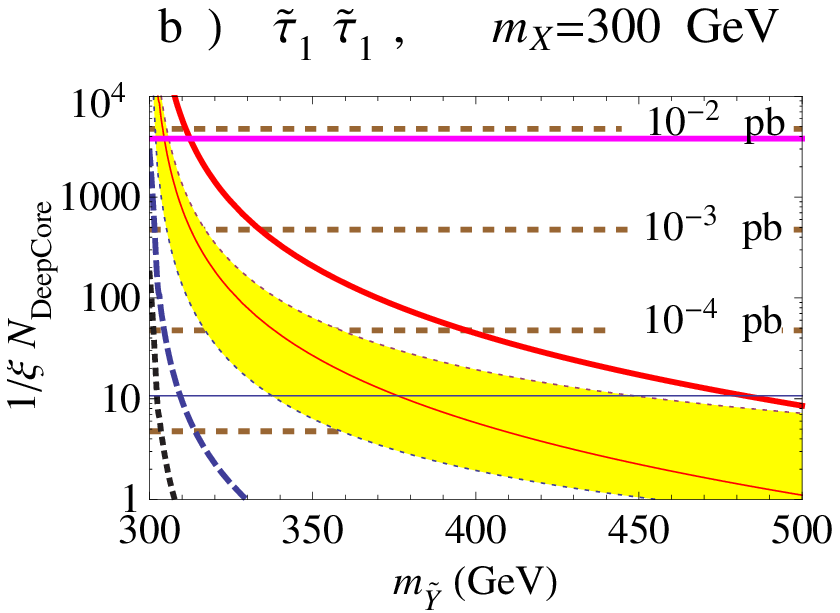}\
\includegraphics[width=53mm, height=40mm]{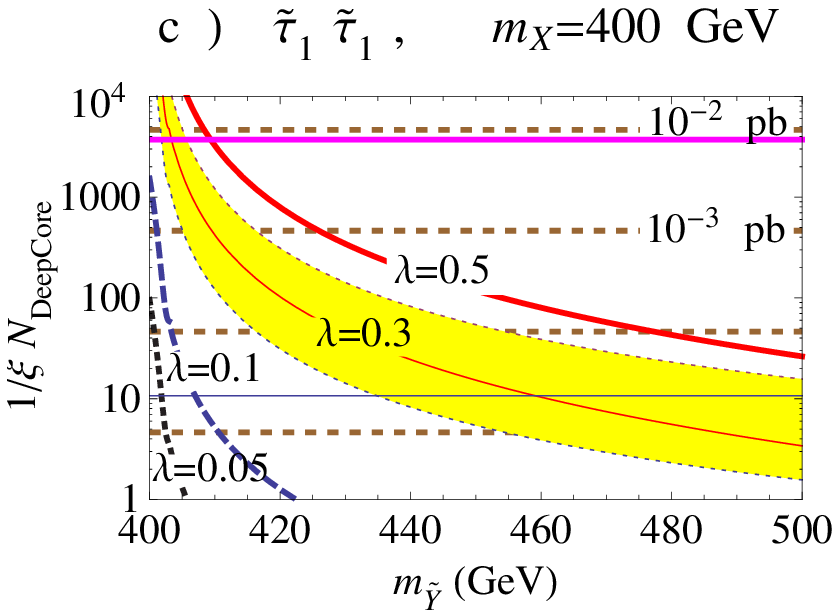}}
	\caption{Similar to Fig.~7 for the $\tilde{\tau}_1^-\tilde{\tau}_1^+$~channel. The thick pink horizontal line represents
the Super-K 90\% C.~L. upper bound on $\sigma_{SD}$.}
\end{figure}

\subsection{Event Rates: Stau and sneutrino channels}

In Figs.~8 and 9 we compare the performances of IceCube and DeepCore, respectively, for the stau~channel. Figs.~10 and 11 show our
results for the sneutrino~channel. 
\begin{figure}[ht!]
	\mbox{\ \includegraphics[width=80mm, height=53mm]{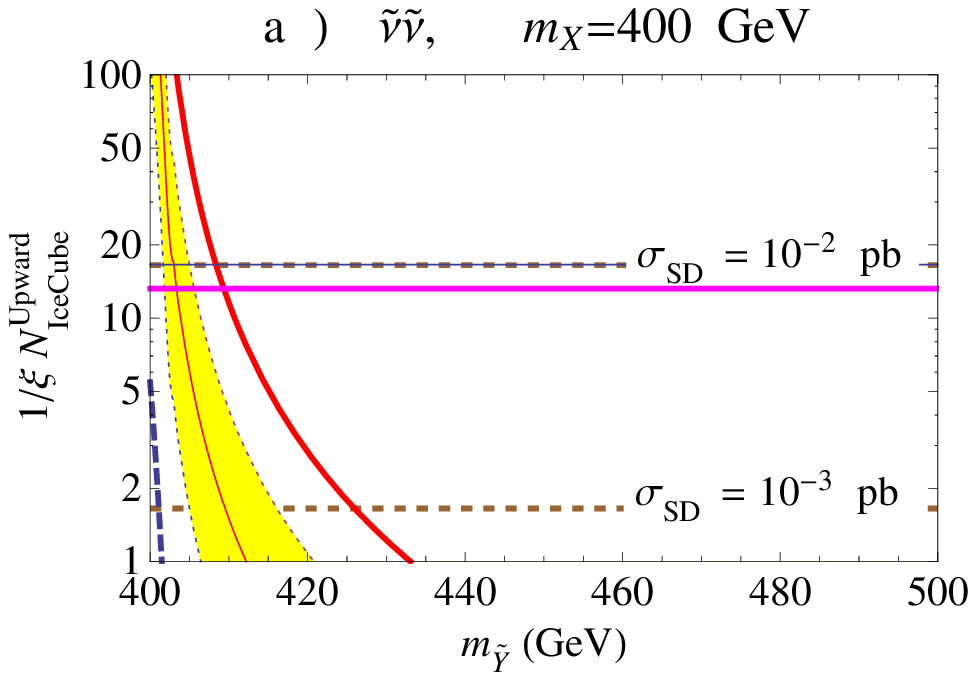}\ \includegraphics[width=80mm, height=53mm]{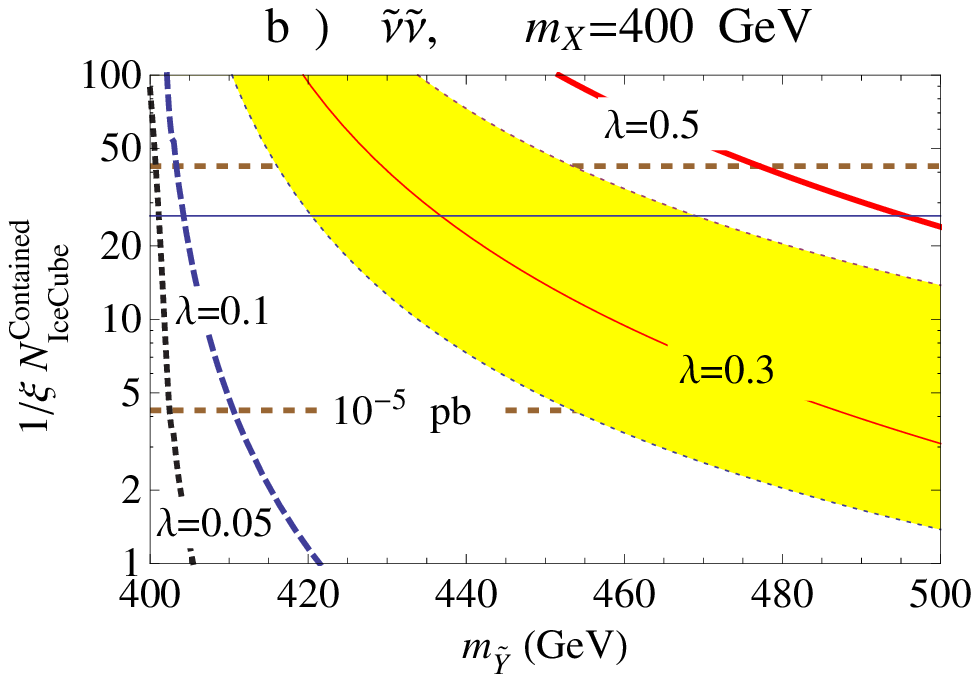}}
	\caption{Similar to Fig.~6 for the $\tilde{\nu}\tilde{\nu}$~channel. IceCube is insensitive to this channel for
\mbox{$m_{X}=150$~GeV, 300~GeV}. The thick pink horizontal line represents the Super-K 90\% C.~L. upper bound on $\sigma_{SD}$.
In a), the solid blue horizontal line is coincident with the dashed brown line representing $\sigma_{SD}=10^{-2}$~pb.}
\end{figure}
\begin{figure}[ht!]
	\mbox{\ \includegraphics[width=53mm, height=40mm]{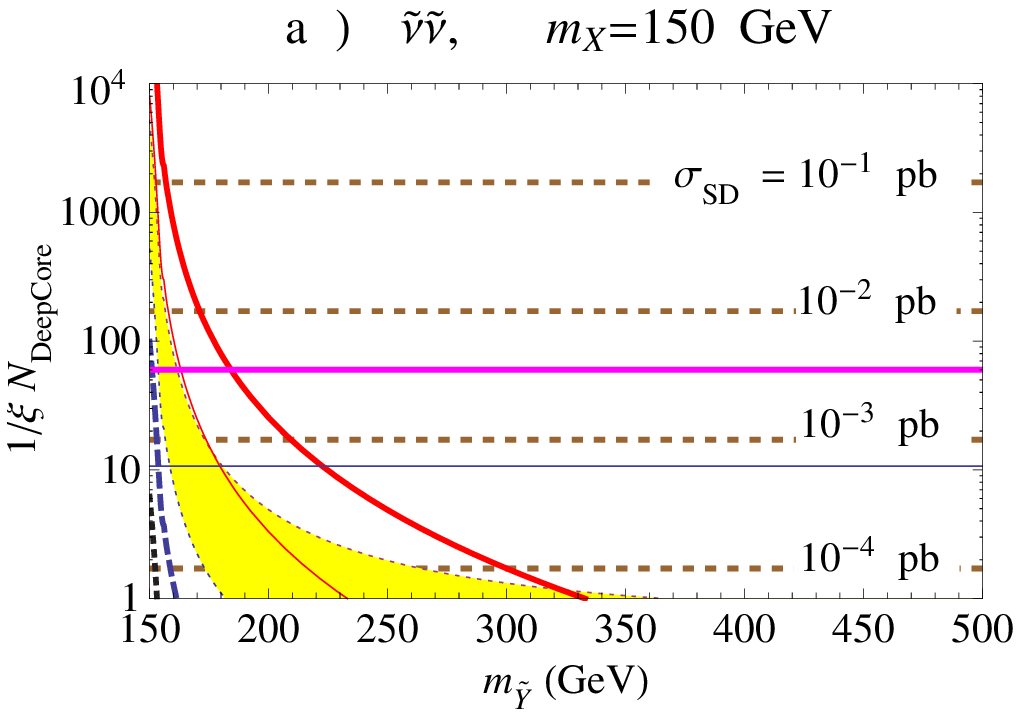}\ \includegraphics[width=53mm, height=40mm]{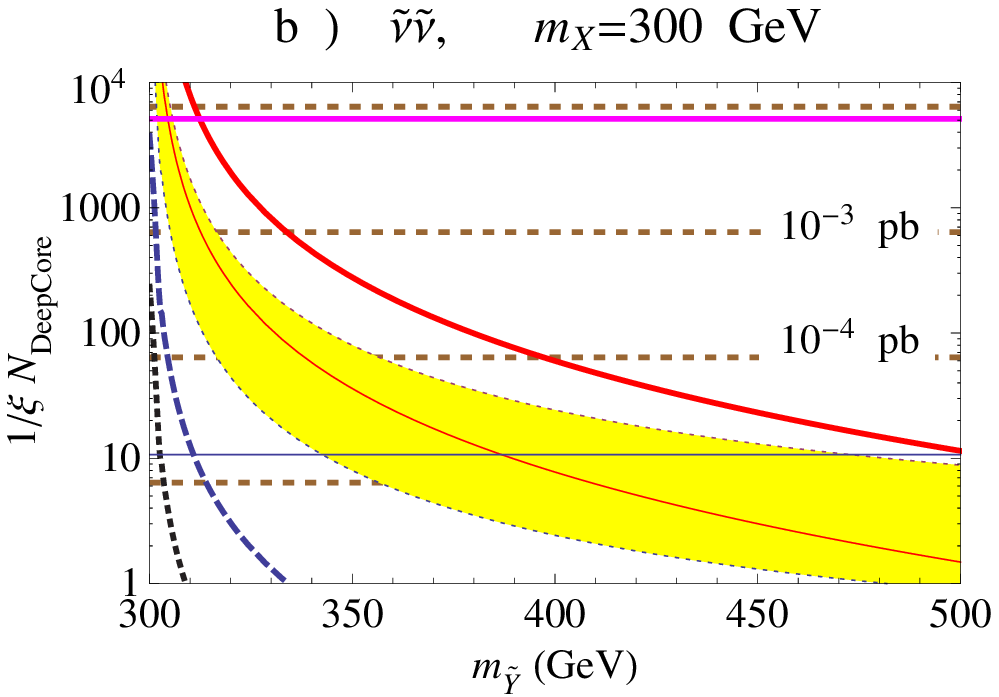}\
\includegraphics[width=53mm, height=40mm]{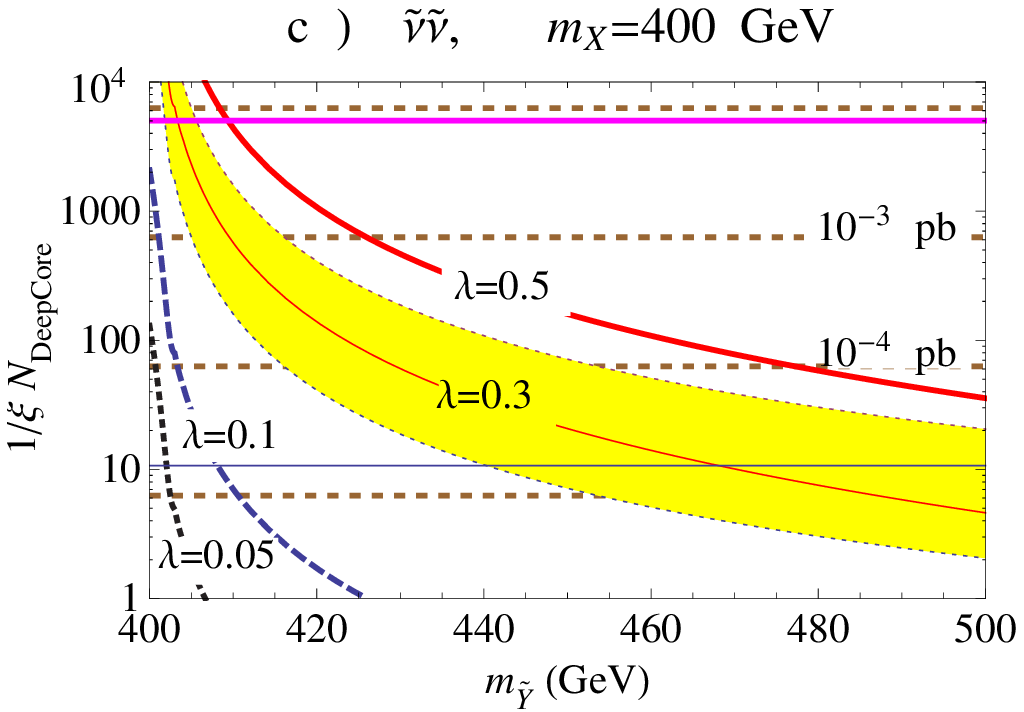}}
	\caption{Similar to Fig.~7 for the $\tilde{\nu}\tilde{\nu}$~channel. The thick pink horizontal line represents the Super-K 90\% C.~L.
upper bound on $\sigma_{SD}$.}
\end{figure}
The 90\% C.~L. upper bound placed by Super-Kamiokande~\cite{Desai:2004pq} on $\sigma_{SD}$ is indicated on
the plots by a thick horizontal pink line. For $m_{X}=150$~GeV this bound is
$\sigma_{SD}^{Max}\simeq3.5\times10^{-3}$~pb and
for $m_{X}=300-400$~GeV, $\sigma_{SD}^{Max}\simeq8\times10^{-3}$~pb. For the $\tau^-\tau^+$~channel
these bounds are off scale and therefore not visible. The advantages of DeepCore in the
observation of neutrino fluxes from low-mass DM are evident:

$\bullet$ For our choice of spectra and energy thresholds,
DM particles with a 150~GeV (300~GeV) mass do not produce observable events at IceCube in the two sparticle channels (sneutrino~channel).
At production, neutrino
spectra for annihilation of $m_{X}=150$~GeV DM particles become negligible above 60~GeV for the stau~channel and
above 40~GeV for the sneutrino~channel. Neutrino spectra originating from $m_{X}=300$~GeV DM annihilation into
sneutrinos are negligible above 80~GeV. Thus, with a threshold energy $E_{thr}=100$~GeV no signal is expected for these channels at IceCube.

$\bullet$ As Fig.~8a shows, even if observation of upward events at IceCube is possible in the stau~channel for
\mbox{$m_{X}=300$~GeV}, the region of parameter space
allowed is very narrow. It is represented by the strip between the thin blue and thick pink horizontal lines. Similarly,
Fig.~10a shows that a 3$\sigma$~discovery in upward events for $m_{X}=400$~GeV in the sneutrino~channel is incompatible with
Super-K data, as the thin blue line is above the thick pink line. Figs.~8c and 10b show that, as usual, the contained events
offer better detection prospects at IceCube: assuming $\lambda=0.5$, \mbox{3$\sigma$~detection} is obtained with a connector
mass within 31\% of $m_{X}=300$~GeV in the stau~channel, and within 24\% of $m_{X}=400$~GeV in the sneutrino~channel.\medskip

The performance of DeepCore is much less sensitive to the annihilation channel than
IceCube, because the lower energy
threshold allows integration of much of the neutrino spectra independent of its shape or features.

$\bullet$ Even for values of $m_{X}$ that do not allow observation of events at Icecube, DeepCore is able to
observe events at 3$\sigma$ in 5~yrs for reasonable values of $\lambda$ and $m_{\tilde{Y}}$. However, the region of parameter
space for which DeepCore is sensitive to the $m_{X}=150$~GeV sparticle channels is very
narrow, as can be seen in Figs.~9a and 11a.

$\bullet$ The advantages of DeepCore over IceCube are not
substantial for heavy-DM masses in the stau~channel, and the detector fares just a little better in the heavy-DM sneutrino
channel. Fig.~9c (11c) shows that 3$\sigma$~detection of $m_{X}=400$~GeV DM in the
stau (sneutrino) channel can be obtained with a range of connector masses extended by a mere 3\% (18\%)
with respect to IceCube.

$\bullet$ The yellow bands represent the region of $(m_{\tilde{Y}},\lambda)$-parameter
space such that
$0.1~{\rm pb}\leq\sigma_{X\bar{X}\rightarrow\tilde{f}_{i}\bar{\tilde{f}}_{i}} v\leq 1~{\rm pb}$,
assuming that $\lambda'_i=\lambda$ and $m_{Y^{lep.}}=m_{\tilde Y}$.\bigskip

It has been recently pointed out~\cite{Mandal:2009yk} that DeepCore should be able to identify contained ``cascade''
events, which
originate from CC~interactions of electron and tau neutrinos inside the detector, and from NC~interactions of
neutrinos of all
three flavors. For such events the signal is enhanced with respect to the background at
energies $\gtrsim 40$~GeV, since the signal is predominantly
due to the CC~interactions. Specifically, at such energies the flux of atmospheric
$\nu_\mu$ is from one to a few orders of magnitude larger than the flux of $\nu_e$ and
$\nu_\tau$~\cite{Stanev:1999ki}; thus,
$\nu_\mu$ give the dominant contribution to the background through NC~interactions,
which are weaker than CC~interactions. This could provide an additional method of detection
for neutrinos. However, as noted in~\cite{Resconi:2008fe,Middell}, the angular sensitivity
for cascades is $\sim 50^{\circ}$, which does not permit tracking of the Sun
with the desired accuracy. Consequently, we do not consider cascade events as a potential signal.

\section{Conclusions}

We investigated the prospects for indirect detection of fermion WIMPless DM at IceCube and DeepCore.
We considered a hidden sector Majorana DM particle of mass $m_{X}$ that couples through Yukawa couplings in the
superpotential to a
connector of mass $m_{Y}$ and visible sector particles.  These models are especially interesting in the context
of IceCube/DeepCore
because they exhibit only SD nuclear scattering, for which IceCube/DeepCore is expected to soon provide the
greatest experimental sensitivity for $m_X \sim {\cal O} (\textrm{few } 100~{\rm GeV})$.

We focused our attention on DM annihilation to taus, staus and sneutrinos. In order to be captured by the Sun
the DM particle needs to couple also to up and down quarks and first generation squarks through the superfield.
Annihilation to light fermions does not
produce an observable signal because it is chirality suppressed. As for annihilation into squarks, in
the GMSB-inspired WIMPless scenario, squarks are generally very heavy and the resulting neutrino spectra are strongly dependent on
the features of the sparticle spectrum. Therefore, we assumed
that the dominant channels for
annihilation are exclusively leptonic. We
assumed that if the DM candidate annihilates to SM particles, these are predominantly taus. If it annihilates
to supersymmetric
particles, these belong to lepton superfields and are the NNLSP, with the lightest neutralino being the NLSP.
In the cases of stau and sneutrino
annihilation channels we
focused our attention on the distinct situations where either the DM candidate couples predominantly to the
third family (staus), or it couples
with equal strength to degenerate sneutrinos of all three families. We believe this allows us to consider a
range of interesting possibilities.

We propagated the neutrino spectra originated by DM annihilation at the center of the Sun through the
solar medium. We took into account
energy losses due to NC~interactions, CC~interactions, and tau~regeneration. We also implemented neutrino
oscillations in the propagation
through the Sun.

In order to calculate the event rate at the IceCube detector, we considered upward and contained events,
taking into account muon energy
losses due to ionization,
bremsstrahlung, pair production, and photonuclear effects. The technical details of our analyses
are provided in the appendices.

We found that it is not
possible to obtain 3$\sigma$~detection in upward or contained events at IceCube for the stau and sneutrino channel
for DM masses $m_{X}\lesssim 150$~GeV with 5~yrs of data. Moreover, a 300~GeV DM particle would not produce
observable events for the sparticle channels, unless the Yukawa couplings are large or the connector mass is not more than
about 30\% heavier than the DM mass. Indirect detection of these relatively light DM particles is favored in the tau~channel,
due to the broader neutrino spectra produced in this channel. We quantified the improvement DeepCore
brings to the detection prospects in all channels, especially for low $m_{X}$. In particular,
even in the cases mentioned above, where the steep decrease in flux below the detector energy threshold makes
signal detection highly unlikely at IceCube, such
limitations do not apply to DeepCore. While the performance
of the IceCube detector varies significantly between different channels and different DM masses, thus requiring
a case by case analysis, we
showed that DeepCore can comfortably produce a 5~yr 3$\sigma$~detection for any analyzed channel, without
strong dependence on the Yukawa
couplings or connector masses. We thus find good prospects for probing models of Majorana fermionic WIMPless
DM at IceCube, including the
DeepCore extension.


\section*{Acknowledgments}

We thank A.~Erkoca, J.~Feng, F.~Halzen, I.~Mocioiu, B.~Morse, M.~H.~Reno and I.~Sarcevic for useful discussions, S. Mrenna and
P. Skands for help with Pythia, D.~Besson, T.~DeYoung, D.~Grant, and C.~Rott for details about DeepCore/IceCube,
and especially Y. Gao for many
discussions and inputs. This work was
supported by the DoE under Grant Nos. DE-FG02-95ER40896 and DE-FG02-04ER41308, by the NSF under
Grant No. PHY-0544278, and by the Wisconsin Alumni Research Foundation.

\appendix

\section{Neutrino Propagation}

We describe the procedure involving the propagation of neutrinos from the
center of the Sun to the Earth, and the detection at IceCube and DeepCore.

Once the neutrinos are produced at the center of the Sun, they need to be propagated through the solar medium,
travel to the Earth and be detected at the neutrino telescope. The
appropriate formalism involves the density matrix for the neutrino
spectra in the flavor basis~\cite{Cirelli:2005gh}. We call this
$\mathbf{\rho}$, and indicate matrices in boldface.

Neutrinos of energy $E_{\nu}$ can be propagated from a point $r$ to $r+dr$ inside the Sun by solving
the Heisenberg equation,
\begin{equation}
    \frac{d\mathbf{\rho}(E_{\nu})}{dr}=-i
    [\mathbf{H}(E_{\nu}),\mathbf{\rho}(E_{\nu})]+\left.\frac{d\mathbf{\rho}(E_{\nu})}{dr}\right|_{NC}+
    \left.\frac{d\mathbf{\rho}(E_{\nu})}{dr}\right|_{CC}+\left.\frac{d\mathbf{\rho}(E_{\nu})}{dr}\right|_{in}\,,\label{HeisEq}
\end{equation}
where $\mathbf{H}$ is the Hamiltonian for neutrino
oscillations in matter, the term indicated by $in$ is the
injection spectrum at the center of the Sun and the
other two terms represent the matter effects due to NC- and CC-interactions. The Hamiltonian is
\begin{equation}
    \mathbf{H}=\frac{1}{2E_{\nu}}\mathbf{U} \textrm{diag}(0,\Delta m^{2}_{21},\Delta
    m^{2}_{31})\mathbf{U}^{\dag}+\textrm{diag}(\sqrt{2}G_{F}N_{e},0,0)\,,\label{Hamiltonian}
\end{equation}
\begin{figure}[h!]
 \begin{center}
  \includegraphics[width=90mm, height=60mm]{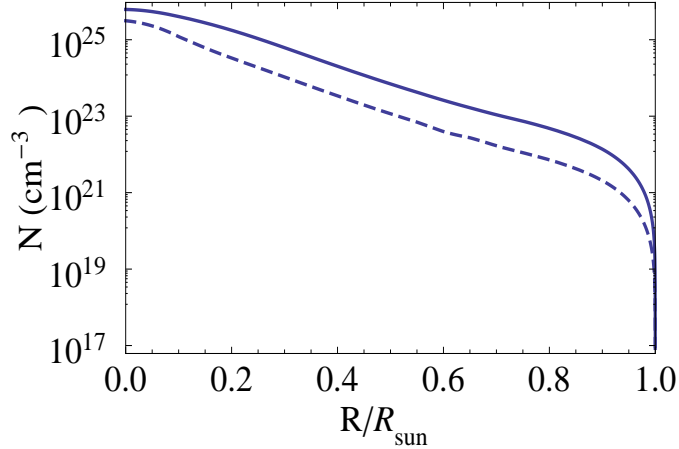}
     \caption{Electron density (solid line) and neutron density (dashed line) as a function of the Sun's radius.}
   \end{center}
\end{figure}
where $\mathbf{U}$ is the neutrino mixing matrix as given in~\cite{Amsler:2008zzb}, $N_{e}$
is the radius-dependent density of electrons inside the Sun,
$G_{F}$ is the Fermi constant, $E_{\nu}$ is the energy of the
incoming neutrino, and $\Delta m_{21}^2$ and $\Delta m_{31}^2$ are the
neutrino mass-squared differences, $\Delta m_{ij}^2=m_i^2-m_j^2$. We use the following
values: $\Delta m_{21}^2=8.1\times10^{-5}$ eV$^2$, $\Delta
m_{31}^2=2.2\times10^{-3}$ eV$^2$, $\theta_{12}=33.2^{\circ}$,
$\theta_{23}=45^{\circ}$, $\theta_{13}=0$. It has been shown in~\cite{Cirelli:2005gh,Barger:2007xf} that the
oscillation results are not significantly changed if $\theta_{13}$ is small but nonzero.
The density profile of the
Sun is shown in Fig.~12~\cite{Bahcall:2000nu}.

The injection term is
\begin{equation}
    \left.\frac{d\mathbf{\rho}_{ij}}{dr}\right|_{in}(E_{\nu})=\delta(r)\delta_{ij}\frac{dN}{dE_{\nu}}\,.\label{injection}
\end{equation}
It expresses the number of neutrinos and antineutrinos produced in DM-annihilation at
the center of the Sun as a function of energy. It is diagonal in the flavor basis.

In considering the NC and CC terms, we
introduce the following quantity which depends on the deep inelastic
scattering (DIS) total cross sections $\sigma$:
\begin{equation}
    \mathbf{\Gamma}_{NC(CC)}(E_{\nu},E')=N_{p}(r)\textrm{diag}[\sigma(\nu_{l}p\rightarrow\nu'_{l}(l)+\textrm{any})]
    +    N_{n}(r)\textrm{diag}[\sigma(\nu_{l}n\rightarrow\nu'_{l}(l)+\textrm{any})]\,,\label{gammamatr}
\end{equation}
where $N_{p}(r)(=N_e(r))$ and $N_{n}(r)$ are the proton and neutron
densities inside the Sun, plotted in Fig.~12, $E_{\nu}$ is the incoming neutrino energy, $E'$ the outgoing neutrino
(charged lepton) energy and $l$ labels
flavor. Then, the neutral-current term is given by
\begin{equation}
    \left.\frac{d\mathbf{\rho}(E_{\nu})}{dr}\right|_{NC}=-\mathbf{\rho}(E_{\nu})
    \int_{0}^{E_{\nu}}\frac{d\mathbf{\Gamma}_{NC}}{dE'_{\nu}}(E_{\nu},E'_{\nu})dE'_{\nu}+
    \int_{E_{\nu}}^{\infty}\frac{d\mathbf{\Gamma}_{NC}}{dE_{\nu}}(E'_{\nu},E_{\nu})\mathbf{\rho}(E'_{\nu})dE'_{\nu}\,.
    \label{NC}
\end{equation}

The CC term is defined in a similar way but it is
more complicated for two reasons. First, the CC-DIS cross
sections are not the same for all flavors of neutrinos. As a
matter of fact the $\nu_{\tau}$ cross sections are suppressed near threshold
by the kinematical effects of $m_{\tau}$. Second,
one needs to take into account the effects of tau regeneration that couple the
propagation of different flavors (different elements of the
density matrix) and elements of the neutrino and antineutrino
density matrices.

Tau regeneration is an important effect that leads to a
reinjection of the neutrinos produced by the decay of
the taus that are CC created by neutrinos of higher energies. The
taus produced by energetic neutrinos undergoing CC
interactions can decay promptly through various channels, for example,
$\tau^{-}\rightarrow\nu_{\tau}+\textrm{any}$, $\tau^{-}\rightarrow
e^{-}\bar{\nu}_{e}\nu_{\tau}$ and $\tau^{-}\rightarrow
\mu^{-}\bar{\nu}_{\mu}\nu_{\tau}$, and similarly for the
antiparticles. These processes provide additional sources of
energetic neutrinos that reenter the flux with lower energies. The
probabilities of reinjection are encoded in four functions
$f_{\nu_{\tau}\rightarrow\nu_{\tau}}(u)$,
$f_{\bar{\nu}_{\tau}\rightarrow\bar{\nu}_{\tau}}(u)$,
$f_{\nu_{\tau}\rightarrow\bar{\nu}_{e,\mu}}(u)$,
$f_{\bar{\nu}_{\tau}\rightarrow\nu_{e,\mu}}(u)$, which depend on
the branching ratios of the above channels and on $u\equiv
E_{\nu}^{out}/E_{\nu}^{in}$, where $E_{\nu}^{in}$ is the energy of
the tau neutrino undergoing CC scattering, and $E_{\nu}^{out}$ is
the energy of the lower energy neutrinos produced by tau decay~\cite{Cirelli:2005gh}.
The charged-current contribution to the
Heisenberg equation is therefore,
\begin{eqnarray}
    \left.\frac{d\mathbf{\rho}(E_{\nu})}{dr}\right|_{CC}&=&-\frac{\{\mathbf{\Gamma}_{CC},\mathbf{\rho}\}}{2}+
    \int_{E_{\nu}}^{\infty}\frac{dE_{\nu}^{in}}{E_{\nu}^{in}}
    \left[\mathbf{\Pi}_{\tau}\rho_{\tau\tau}(E_{\nu}^{in})\Gamma^{\tau}_{CC}(E^{in}_{\nu})
    f_{\nu_{\tau}\rightarrow\nu_{\tau}}\left(\frac{E_{\nu}}{E_{\nu}^{in}}\right)\right.\nonumber\\
     & &+\left.\mathbf{\Pi}_{e,\mu}\bar{\rho}_{\tau\tau}(E_{\nu}^{in})\bar{\Gamma}^{\tau}_{CC}(E^{in}_{\nu})
    f_{\bar{\nu}_{\tau}\rightarrow\nu_{e,\mu}}\left(\frac{E_{\nu}}{E_{\nu}^{in}}\right)\right]\,,\label{rhoCC}\\
    \left.\frac{d\mathbf{\bar{\rho}}(E_{\nu})}{dr}\right|_{CC}&=&-\frac{\{\mathbf{\bar{\Gamma}}_{CC},\mathbf{\bar{\rho}}\}}{2}+
    \int_{E_{\nu}}^{\infty}\frac{dE_{\nu}^{in}}{E_{\nu}^{in}}
    \left[\mathbf{\Pi}_{\tau}\bar{\rho}_{\tau\tau}(E_{\nu}^{in})\bar{\Gamma}^{\tau}_{CC}(E^{in}_{\nu})
    f_{\bar{\nu}_{\tau}\rightarrow\bar{\nu}_{\tau}}\left(\frac{E_{\nu}}{E_{\nu}^{in}}\right)\right.\nonumber\\
     & &+\left.\mathbf{\Pi}_{e,\mu}\rho_{\tau\tau}(E_{\nu}^{in})\Gamma^{\tau}_{CC}(E^{in}_{\nu})
    f_{\nu_{\tau}\rightarrow\bar{\nu}_{e,\mu}}\left(\frac{E_{\nu}}{E_{\nu}^{in}}\right)\right]\,,\label{rhobarCC}
\end{eqnarray}
where $\mathbf{\Pi}_{e}=\textrm{diag}(1,0,0)$ are projectors, and
similar expressions apply to the other flavors. As is clear
from the last term on the right-hand side of Eqs.~(\ref{rhoCC}) and (\ref{rhobarCC}), tau-regeneration effects
couple the two sets of equations. The regeneration probability functions are given by~\cite{Barger:2007xf},
\begin{eqnarray}
 f_{\nu_{\tau}\rightarrow\nu_{\tau}}(u)&=&N\int_{u}^{1}\frac{dz}{z}\left(1+\frac{z^{2}}{5}\right)
 \sum_{i=1}^{6}\textrm{Br}_{i}\left(g_{0i}\left(\frac{u}{z}\right)-g_{1i}\left(\frac{u}{z}\right)\right)\,,\label{f1}\\
f_{\bar{\nu}_{\tau}\rightarrow\bar{\nu}_{\tau}}(u)&=&N\int_{u}^{1}\frac{dz}{z}\left(\frac{1}{5}+z^{2}\right)
\sum_{i=1}^{6}\textrm{Br}_{i}\left(g_{0i}\left(\frac{u}{z}\right)-g_{1i}\left(\frac{u}{z}\right)\right)\,,\label{f2}\\
f_{\nu_{\tau}\rightarrow\bar{\nu}_{e,\mu}}(u)&=&N\int_{u}^{1}\frac{dz}{z}\left(1+\frac{z^{2}}{5}\right)0.18
\left(\tilde{g}_{0}\left(\frac{u}{z}\right)-\tilde{g}_{1}\left(\frac{u}{z}\right)\right)\,,\label{f3}\\
f_{\bar{\nu}_{\tau}\rightarrow\nu_{e,\mu}}(u)&=&N\int_{u}^{1}\frac{dz}{z}\left(\frac{1}{5}+z^{2}\right)0.18
\left(\tilde{g}_{0}\left(\frac{u}{z}\right)-\tilde{g}_{1}\left(\frac{u}{z}\right)\right)\,,\label{f4}
\end{eqnarray}
where $z=E_{\tau}/E_{\nu}^{in}$ and $N$ normalizes the equations so that
their integral is either 1 ($\nu_{\tau}\rightarrow\nu_{\tau}$ and
$\bar{\nu}_{\tau}\rightarrow\bar{\nu}_{\tau}$), or 0.18
($\nu_{\tau}\rightarrow\bar{\nu}_{e,\mu}$ and
$\bar{\nu}_{\tau}\rightarrow\nu_{e,\mu}$). The functions $g_{0i}$
and $g_{1i}$ are, respectively, the unpolarized and polarized energy spectra of the $\tau$~neutrinos originating
from the taus in the fragmentation frame, for each final state. The functions $\tilde{g}_{0}$ and $\tilde{g}_{1}$
are the unpolarized and polarized energy spectra of the ${\bar \nu}_{e,\mu}$ from decay of the $\tau$. The explicit
forms of $g_{0i}$ and $g_{1i}$ are given in Table~I of~\cite{Dutta:2000jv}, and those of $\tilde{g}_{0}$
and $\tilde{g}_{1}$ are given in Ref.~\cite{Lipari:1993hd}.
The branching fractions
Br$_{i}$ refer to the six possible final states:
$\tau^{-}\rightarrow e^{-}\bar{\nu}_{e}\nu_{\tau}$
(Br$_{1}=$0.18), $\tau^{-}\rightarrow
\mu^{-}\bar{\nu}_{\mu}\nu_{\tau}$ (Br$_{2}=$0.18),
$\tau^{-}\rightarrow \pi^{-}\nu_{\tau}$ (Br$_{3}=$0.12),
$\tau^{-}\rightarrow a_{1}\nu_{\tau}$ (Br$_{4}=$0.13),
$\tau^{-}\rightarrow \rho\nu_{\tau}$ (Br$_{5}=$0.26),
$\tau^{-}\rightarrow \nu_{\tau}+\textrm{any}$ (Br$_{6}=$0.13).

Once the neutrinos reach the surface of the Sun, the propagation to the Earth is obtained by the following averaging
procedure: we rotate the density matrix to
the mass basis; drop the off-diagonal terms and rotate it
back to the flavor basis. With our choice of neutrino parameters, the averaging should wash out any observable modulation.

\section{Muon Rates}

\subsection*{Upward Events at IceCube}

When a muon generated via CC~interactions travels through the rock and ice
beneath the detector it loses energy due to ionization,
bremsstrahlung, pair production, and photonuclear effects~\cite{Lipari:1991ut}.
The average energy loss of the muons that travel a distance $dz$ in a medium of density $\rho_{med}$ is given by:
\begin{equation}
 \left\langle\frac{dE}{dz}\right\rangle=-(\alpha+\beta(E) E)\rho_{med}(z)\,,\label{energyloss}
\end{equation}
 where $\alpha=3.0\times 10^{-3}$~GeV~cm$^2/$g is related to ionization, while $\beta(E)$ takes into
 account bremsstrahlung,
 pair production, and photonuclear effects. We take
 $\beta=3.0\times 10^{-6}$~cm$^2/$g and $\rho_{med}=\rho_{ice}=0.92$~g/cm$^3$. The results from Muon
 Monte Carlo~\cite{Chirkin:2004hz} are reproduced by choosing these values of $\alpha$ and $\beta$~\cite{YuGao}.
 Equation~(\ref{energyloss})
 can then be easily solved to obtain the final energy $E_{\mu f}$, given initial energy $E_{\mu i}$:
\begin{equation}
 E_{\mu f}=-\frac{\alpha}{\beta}+e^{-\beta\rho_{ice} z}\left(E_{\mu i}+\frac{\alpha}{\beta}\right)\,.\label{relatEiEf}
\end{equation}
The average range covered by the muon between energies $E_{\mu i}$, $E_{\mu f}$ is then,
\begin{equation}
 R_{\mu}(E_{\mu i},E_{\mu f})=\frac{1}{\beta\rho_{ice}}\ln\left(\frac{\alpha+\beta E_{\mu i}}{\alpha+\beta
 E_{\mu f}}\right)\,.\label{murange}
\end{equation}

Thus, the muon flux at the detector is obtained by a convolution of the following:
the probability of the incoming neutrino to CC~scatter with a nucleus in ice;
the average range over which energy losses force the muon energy
below the detector threshold; the muon probability of surviving its own decay length.
This last effect can be parametrized by the survival probability,
\begin{equation}
 P_{sur}(E_{\mu i},E_{\mu f})=\left[\frac{E_{\mu f}(\alpha+\beta E_{\mu i})}
{E_{\mu i}(\alpha+\beta E_{\mu f})}\right]^{\frac{m_{\mu}}{c\tau\alpha\rho_{ice}}}\,,\label{survP1}
\end{equation}
which is a solution to the differential equation,
\begin{equation}
 \frac{dP_{sur}}{dE_{\mu f}}=\frac{P_{sur}}{E_{\mu f}c\tau\rho_{ice}(\alpha+\beta E_{\mu f})/m_{\mu}}\,,\label{survP2}
\end{equation}
where $\tau$ is the muon lifetime and $m_{\mu}$ its mass. Folding these effects together gives the spectrum of
muon events,
\begin{eqnarray}
 \frac{d\Phi_{\mu}}{dE_{\mu f}}&=&\int_{0}^{R_{\mu}(m_{X},E_{\mu f})}dz\, e^{\beta\rho_{ice} z}P_{sur}(E_{\mu i}
 (E_{\mu f},z),E_{\mu f})\nonumber\\
 &\times&\int_{E_{\mu i}(E_{\mu f},z)}^{m_{X}}dE_{\nu}
 \left[\frac{d\Phi_{\nu}}{dE_{\nu}}\left(\frac{d\sigma^{CC}_{\nu p}}{dE_{\mu i}}(E_{\nu})\rho_p
 +\frac{d\sigma^{CC}_{\nu n}}
 {dE_{\mu i}}(E_{\nu})\rho_n\right)\right.\nonumber\\
 &+&\left.(\textrm{a corresponding contribution for }\bar{\nu})\right]\,,\label{nuflux}
\end{eqnarray}
where $E_{\mu i}(E_{\mu f},z)$ is obtained by inverting Eq.~(\ref{relatEiEf}), $d\sigma^{CC}/dE_{\mu}(E_{\nu})$
are the differential CC-cross sections to protons ($\nu p$) and neutrons ($\nu n$),
$\rho_{p}\sim5/9\textrm{ }N_{A}$ cm$^{-3}$
and $\rho_{n}\sim4/9\textrm{ }N_{A}$ cm$^{-3}$ are the number densities of nucleons in ice
expressed in terms of Avogadro's number~$N_{A}$, and $d\Phi_{\nu}/dE_{\nu}$ is the neutrino spectrum at
Earth, depicted in our
specific cases in the third column of \mbox{Figs.~3, 4 and 5.}

The event rate for upward events is obtained by
convolving Eq.~(\ref{nuflux}) with the muon
\textit{effective area} of the detector, $A_{eff}(E_{\mu f})R(\cos\theta)$, which is constituted by a zenith
angle-independent part, shown in Fig.~13a, and a factor $R(\cos\theta)=0.92-0.45\cos\theta$ that accounts
for the rock bed beneath the ice~\cite{GonzalezGarcia:2009jc}. We take the average of the effective area over
the time of the year that the Sun
spends below the horizon, namely between the March and September equinoxes. We define the zenith angle
$\theta_{z}$ at the
South Pole to be the angle centered at the detector with $\theta_{z}=0^{\circ}$ indicating the vertical direction in the
sky. $\theta_{z}$ can be parametrized in terms of the time of the year $f_{y}$ (where $f_y=0,1/2$ correspond to the March
and September equinoxes, respectively), and the tilt of the Earth axis
with respect to the perpendicular to the ecliptic plane, $\theta_{t}=23^{\circ} 26'$:
\begin{equation}
 \theta_z(f_y)=\frac{\pi}{2}+\theta_{t}\sin(2\pi f_{y})\,.\label{thetazeta}
\end{equation}
The event rate reads,
\begin{equation}
    N_{events}^{Up}=\xi\Gamma_{eq}\int_{0}^{m_{X}}\frac{d\Phi_{\mu}}{dE_{\mu f}}\langle A_{eff}(E_{\mu f})
    R(\cos\theta_z)\rangle dE_{\mu f},\label{Nevents}
\end{equation}
where $\xi$ and $\Gamma_{eq}$ are the quantities introduced in Sec.~3.1. $\langle A_{eff}(E_{\mu f})
R(\cos\theta_z)\rangle$
is the average over the portion
of the solid angle ($\theta_{z},\phi$) that corresponds to the time the Sun spends below the horizon.

\begin{figure}[ht!]
	\mbox{\ \includegraphics[width=75mm, height=53mm]{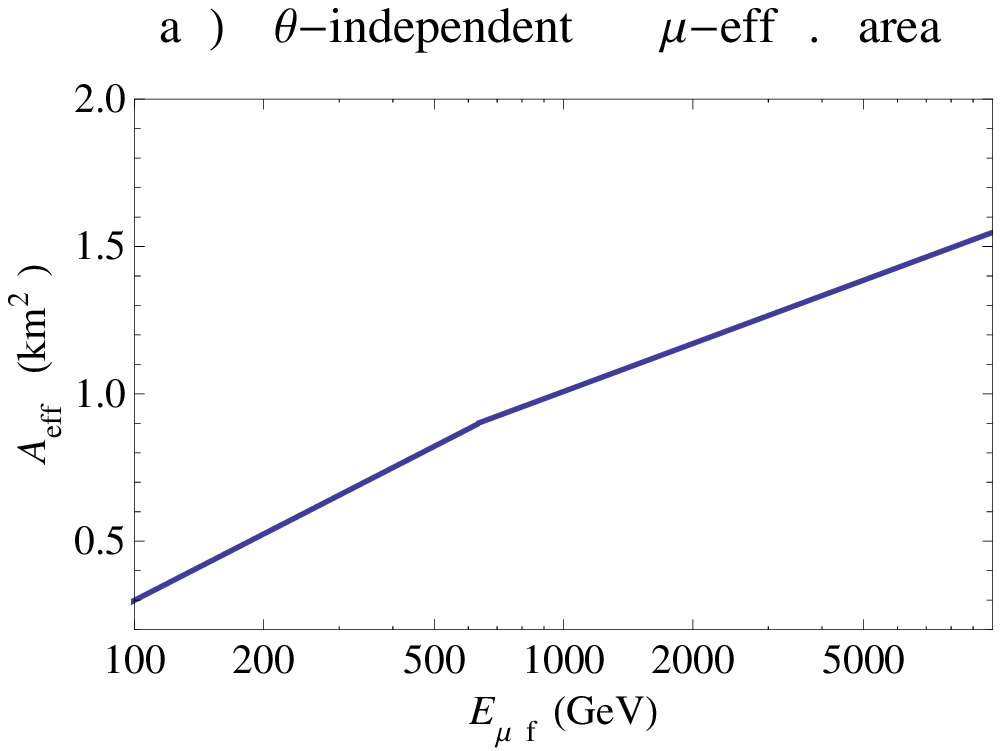}\ \includegraphics[width=80mm, height=53mm]{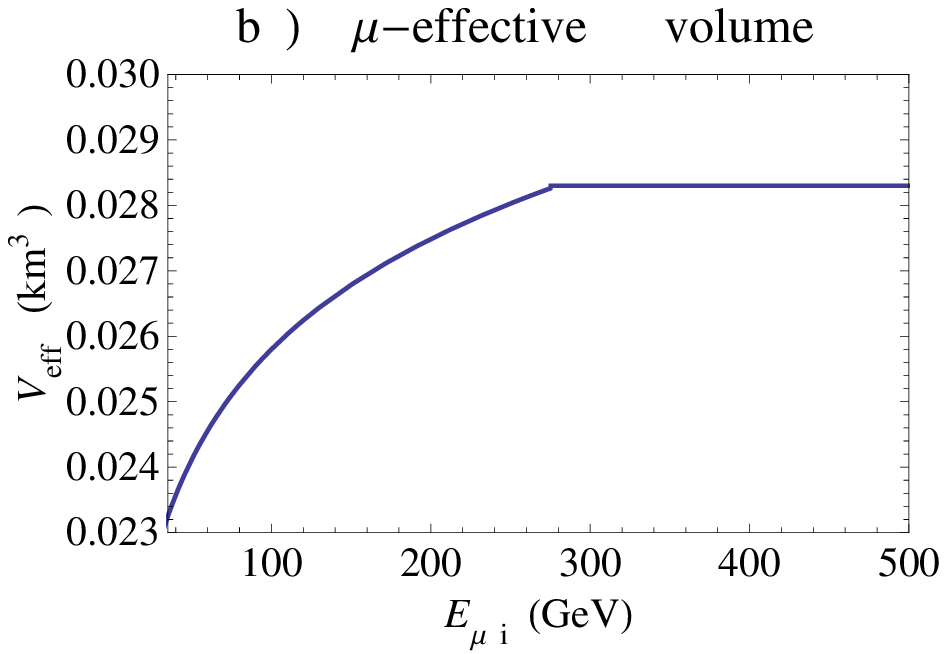}}
	\caption{a)~The IceCube muon effective area with zenith angle dependence factored out.
b)~Muon effective volume for DeepCore.}
\end{figure}

\newpage

\subsection*{Contained Events at IceCube}

For neutrinos that interact within the detector volume, the resulting muons do not lose an appreciable fraction of their
energies. Following notation similar to the previous subsection (except that we identify $E_{\mu}\equiv E_{\mu i}$
since the
muons do not propagate) the
muon flux for contained events is given by
\begin{equation}
    \frac{d\Phi_{\mu}}{dE_{\mu}}=L\int_{E_{\mu}}^{m_{X}}dE_{\nu}
 \left[\frac{d\Phi_{\nu}}{dE_{\nu}}\left(\frac{d\sigma^{CC}_{\nu p}}{dE_{\mu}}(E_{\nu})\rho_p
 +\frac{d\sigma^{CC}_{\nu n}}
 {dE_{\mu}}(E_{\nu})\rho_n\right)
 +(\nu\rightarrow\bar{\nu})\right]\,,\label{nufluxC}
\end{equation}
where $L\sim 1$~km is the size of the IceCube detector. The event rate reads
\begin{equation}
    N_{events}^{C}=\xi\Gamma_{eq}\int_{E_{thr}}^{m_{X}}\frac{d\Phi_{\mu}}{dE_{\mu}}(1\textrm{ km}^2)dE_{\mu}\,,
    \label{NeventsC}
\end{equation}
where $E_{thr}=100$ GeV is the energy threshold of the IceCube detector. As mentioned above, we only
consider events observed
between the March and September equinoxes.

\subsection*{Events at DeepCore}

The great advantage of DeepCore with respect to IceCube is that the outer instrumented volume of IceCube
will serve as a veto
to atmospheric muon events up to one part in $10^6$~\cite{Resconi:2008fe}, so that data can be collected throughout the
year, i.e., even when the Sun is above the horizon.
The rate for contained events at DeepCore can be calculated by convolving Eq.~(\ref{nufluxC}) with the muon
\textit{effective volume} $V_{eff} (E_{\mu i})$. In the most optimistic estimates the effective volume is
constant for muon energies above $\sim300$~GeV, and drops significantly at lower
energies~\cite{DeYoung}. As explained in Sec.~3.3, for DeepCore we
consider the interval $E_{\mu i}> E_{min}=35$~GeV. We find that in this interval the effective volume
in km$^3$ can be parametrized by
\begin{equation}
 V_{eff}(E_{\mu i})=(0.0056\log E_{\mu i}+0.0146) \Theta(275-E_{\mu i})+0.0283 \Theta(E_{\mu i}-275)\,,\label{effvol}
\end{equation}
where $\Theta$ is the Heaviside step function and $E_{\mu i}$ is in GeV. The effective volume is
plotted in Fig.~13b.

After convolution one gets
\begin{equation}
    N_{events}^{DC}=\xi\Gamma_{eq}\int_{E_{min}}^{m_{X}}\frac{1}{L}\frac{d\Phi_{\mu}}{dE_{\mu}}
    V_{eff}(E_{\mu})dE_{\mu}\,.\label{NeventsDC}
\end{equation}

\section{Atmospheric Background}

The angle-dependent flux of atmospheric neutrinos
$d\Phi_{\nu}^{atm}/(dE_{\nu}d\cos\theta)$ is given in~\cite{Honda:2006qj}. These
neutrinos interact with the medium surrounding the detector and
produce muons that constitute the background. As in  the case of neutrinos from annihilation, the
atmospheric background can be divided in upward and contained
events.

By following the notation introduced in the previous subsections
we can write the angular dependence of the upward background flux for IceCube as
\begin{eqnarray}
 \frac{d\Phi_{\mu}^{atm}}{dE_{\mu f}d\cos\theta_z(f_y)}&=&\int_{0}^{R_{\mu}(E_{Max},E_{\mu f})}dz
 e^{\beta\rho_{ice} z}P_{sur}(E_i(E_{\mu f},z),E_{\mu f})\nonumber\\
 &\times&\int_{E_{i}(E_{\mu f},z)}^{E_{Max}}dE_{\nu}\left[\frac{d\Phi_{\nu}^{atm}}{dE_{\nu}d\cos\theta_z(f_y)}
 \left(\frac{d\sigma^{CC}_{\nu p}}{dE_{i}}(E_{\nu})\rho_p+\frac{d\sigma^{CC}_{\nu
 n}}{dE_{i}}(E_{\nu})\rho_n\right)\right.\nonumber\\
 &+&\left.(\nu\rightarrow\bar{\nu})\right]\,,\label{ATMflux}
\end{eqnarray}
where $E_{Max}\sim10^{4}$~GeV is the maximum energy above which
the rapidly falling atmospheric flux becomes negligible. The angle-dependent flux, Eq.~(\ref{ATMflux}), 
is integrated over
a cone of solid angle $d\Omega=\pi(d\theta)^2$, where the opening angle (half the apex angle) 
$d\theta\sim 1^{\circ}$, consistent
with the IceCube angular sensitivity~\cite{Resconi:2008fe,Erkoca:2009by}. The cone tracks the Sun according to
Eq.~(\ref{thetazeta}), for the fraction of the year over which the Sun is below the horizon. The number of
upward background events for IceCube can be
calculated by folding in the effective area $A_{eff}(E_{\mu f})R(\cos\theta_z)$. We find
$N_{BG}^{Up}\simeq6.1$~yr$^{-1}$ where, again, data is taken only between the March  and September Equinoxes.

Contained events are those obtained by muon~neutrinos undergoing charged-current
interactions within the detector volume, thus the angular dependence of the flux is given by
\begin{eqnarray}
 \frac{d\Phi_{\mu}^{atm}}{dE_{\mu}d\cos\theta_z(f_y)}&=& L\int_{E_{\mu}}^{E_{Max}}dE_{\nu}
 \left(\rho_{p}\frac{d\sigma_{\nu}^{p}(E_{\nu},E_{\mu})}{dE_{\mu}}+\rho_{n}\frac{d\sigma_{\nu}^{n}
 (E_{\nu},E_{\mu})}{dE_{\mu}}+\right)\frac{d\Phi_{\nu}^{atm}}{dE_{\nu}d\cos\theta_z(f_y)}\nonumber\\
 &+&(\nu\rightarrow\bar{\nu})\,.\label{Contained}
\end{eqnarray}
The number of contained background events in a cone of $1^{\circ}$ opening for time of
exposure limited to the period between March and September is obtained, as in Eq.~(\ref{NeventsC}),
by convolving with a constant 1~km$^2$ area. We get $N_{BG}^{Con}\simeq15.6$~yr$^{-1}$.

The number of background events for DeepCore over the whole year, obtained by convolving
with $V_{eff}(E_{\mu i})/L$, is $N_{BG}^{DC}\simeq2.5$~yr$^{-1}$.

\begin{figure}[h!]
 \begin{center}
  \includegraphics[width=90mm, height=60mm]{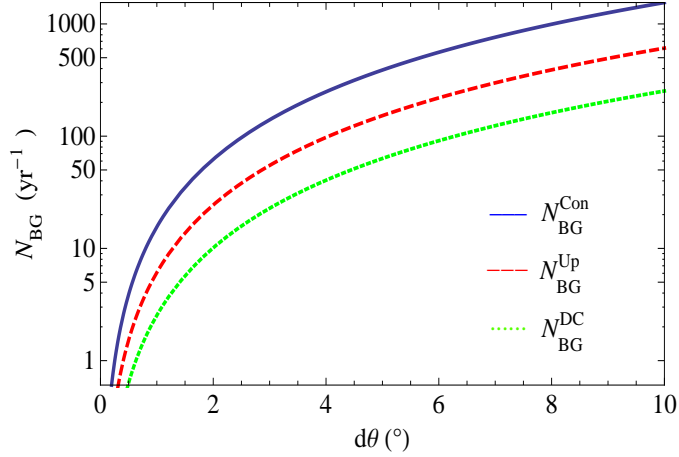}
     \caption{Number of background events per year as a function of the opening angle $d\theta$.}
   \end{center}
\end{figure}

In Fig.~14 we show the dependence of $N_{BG}^{Up}$,
$N_{BG}^{Con}$ and $N_{BG}^{DC}$ on the opening angle $d\theta$. The thin blue horizontal lines shown in Figs.~6-11
represent the 5~yr 3$\sigma$-discovery reach for $d\theta=1^{\circ}$. While the angular resolution is expected to
be best for directions close to the horizon (and the Sun does not stray more than 23$^{\circ}$26' from the horizon at
the South Pole), in case the angular resolution of the IceCube and DeepCore detectors do not meet expectations, the
background event rates can be obtained from Fig.~14. Then, for a 3$\sigma$~detection, \mbox{$N_{\mu}=3\sqrt{N_{BG}}$} 
can be trivially recalculated.

\newpage



\end{document}